\documentclass[twocolumn]{aastex7}
\usepackage{amsmath}
\usepackage{booktabs}

\usepackage[utf8]{inputenc}
\usepackage{verbatim}

\shorttitle{The Politics of Infrastructure}
\shortauthors{Wiechman et al.}

\begin{document}

\title{Politics, Inequality, and  the Robustness of Shared Infrastructure Systems}

\author[orcid=0009-0001-1263-1251]{Adam Wiechman}
\affiliation{Princeton University, Princeton, NJ, 08544, USA}
\email{aw9050@princeton.edu}
\author[orcid=0000-0002-0138-8655]{John M. Anderies}
\affiliation{School of Human Evolution and Social Change and School of Sustainability, Arizona State University, Tempe, AZ 85281, USA}
\email{m.anderies@asu.edu}
\author[orcid=0000-0002-2864-2377]{Margaret Garcia}
\affiliation{School of Sustainable Engineering \& the Built,  Environment, Arizona State University, Tempe, AZ 85281, USA}
\email{M.Garcia@asu.edu}

\keywords{infrastructure $|$ political economy $|$ inequality $|$ robustness $|$ dynamical systems}

\begin{abstract}
Our infrastructure systems enable our well-being by allowing us to move, store, and transform materials and information given considerable social and environmental variation. Critically, this ability is shaped by the degree to which society invests in infrastructure, a fundamentally political question in large public systems. There, infrastructure providers are distinguished from users through political processes, such as elections, and there is considerable heterogeneity among users. Previous political economic models have not taken into account (i) dynamic infrastructures, (ii) dynamic user preferences, and (iii) alternatives to rational actor theory. Meanwhile, engineering often neglects politics. We address these gaps with a general dynamic model of shared infrastructure systems that incorporates theories from political economy, social-ecological systems, and political psychology. We use the model to develop propositions on how multiple characteristics of the political process impact the robustness of shared infrastructure systems to capacity shocks and unequal opportunity for private infrastructure investment. Under user fees, inequality decreases robustness, but taxing private infrastructure use can increase robustness if non-elites have equal political influence. Election cycle periods have a nonlinear effect where increasing them increases robustness up to a point but decreases robustness beyond that point. Further, there is a negative relationship between the ideological sensitivity of candidates and robustness. Overall, the biases of voters and candidates (whether they favor tax increases or decreases) mediate these political-economic effects on robustness because biases may or may not match the reality of system needs (whether system recovery requires tax increases).
\end{abstract}

\section{Introduction}

The fundamental role of infrastructure is to mediate the effect of social and environmental variation on our collective well-being \cite{Anderies2015b,frischmann_infrastructure_2012}. Spatial and temporal variability are endemic to the environmentally derived resources that sustain us - water, flora, fauna, sunlight, wind, etc. - and infrastructure plays a key mediating role in processing such variation (e.g., dams, batteries, and pipelines). However, from the village to the megacity, an often unaccounted for mediating role of infrastructure pertains to social heterogeneity or inequality. In this sense, shared infrastructure both facilitates and emerges from social relations, potentially ``locking in" existing inequalities (e.g., selective transportation projects) or expanding access to basic needs that cannot be met through private means alone (e.g., drinking water) \cite{helmrich_lock-_2023,kanoi_what_2022,pandey_infrastructure_2022}. Given this, \emph{political} considerations abound in what collective well-being goals should be prioritized, what social relationships should be preserved, and what environmental and social variability should be taken into account when making infrastructure investment decisions, as evidenced by the tumultuous federal political process in the early 2020s regarding infrastructure \cite{levin_governance_2022,kanoi_what_2022,anderies_knowledge_2019,Anderies2013b,titolo_bidens_2023}. Yet, many instances of critical infrastructure - drinking water systems, transportation systems, electric grids, legal systems, education systems, etc. - were designed during periods of relative socioeconomic and environmental stability, and a distinguishing feature of the \emph{Anthropocene} is that humanity has pushed Earth, social, economic, and technological systems into a novel period of accelerating and unpredictable change \cite{folke_our_2021,chester_infrastructure_2020,anderies_knowledge_2019}. Within such turbulence, the political nature of infrastructure investment must be incorporated into the dynamics of infrastructure systems \cite{wegrich_infrastructure_2017,Anderies2013b}, but such incorporation remains difficult within the disciplinary nature of most infrastructure scholarship in engineering and economics. We, thus, pursue the following question: \emph{How does the \textbf{political-economic process} of infrastructure investment shape the robustness of shared infrastructure systems to social and environmental variation?}

By robustness, we refer to the preservation of a desired outcome given input variation, or the opposite of sensitivity \cite{Anderies2013a}. In general, the Anthropocene's social and environmental turbulence raises the questions of what attributes of existing shared infrastructure systems can be \emph{preserved} and what new attributes can \emph{emerge} from transitions \cite{folke_our_2021}. Here, we focus on the robustness (preservation) side of the question and on how certain political-economic attributes contribute, or not. Because we do not enter the Anthropocene with a clean slate, any question of emergence must account for the robustness of established systems to identify where new structures might emerge \cite{anderies_knowledge_2019}. 

Our approach leverages theories from political economy, social-ecological systems, and political psychology to build a general model of political-economic feedback in shared infrastructure systems. Disciplinary approaches force scholars to choose one or a few dynamics of interest to simplify mathematical tools or ensure conceptual coherence with established theory in their school of thought \cite{An2012,davidson_simulating_2024}. Studying the overarching feedback processes governing shared infrastructure systems requires, at minimum, consideration of dynamic infrastructure stocks (or ``public capital"), endogenously dynamic policy (e.g., tax levels), inequality in user opportunity to invest in private infrastructure, motivated reasoning in voters and politicians, and political competition. We emphasize that just as it is misleading for engineering models to treat the political-economic process of investment as exogenous, so too, is it misleading for political-economic models to treat infrastructure and the environment as exogenous because they set the \emph{stakes} for politics \cite{wiechman_institutional_2024,Anderies2013a}.

We construct a stylized shared infrastructure system that generalizes beyond particular types (water, transportation, etc.) and define one policy that is dynamically set, the level of taxation. We initialize the shared system to full capacity and conduct a sensitivity analysis to examine the interaction between electoral design, inequality, and assumptions about how voters and candidates learn on the system's ability to withstand a capacity shock and an increase in private infrastructure opportunity for elite users. We perform both a deterministic analysis with a predetermined shock and a stochastic analysis with a given frequency and magnitude of shocks. Our analysis yields multiple propositions related to the political economic determinants of shared infrastructure robustness \cite{selin_progress_2023}. The propositions not only guide hypotheses in future empirical work, but also provide a formal understanding of politics in shared infrastructure systems. We find that (i) private investment opportunity inequality decreases robustness under a user fee system, but (ii) taxing private infrastructure use can increase robustness if tax policy is insulated from elite capture. When voters and candidates face information limitations, (iii) increasing political representation does not guarantee robustness, as voter biases and assumptions about how they learn must be taken into account. Regarding response capacity, there is (iv) a parabolic relationship between election cycle periods and robustness, where increasing the period increases robustness up to a point and further increasing the period decreases robustness. Additionally, (v) increasing the ideological sensitivity of candidates decreases robustness. 

\section{Modeling Shared Infrastructure Political Economics}

\begin{figure*}
    \centering
    \includegraphics[width=0.75\linewidth]{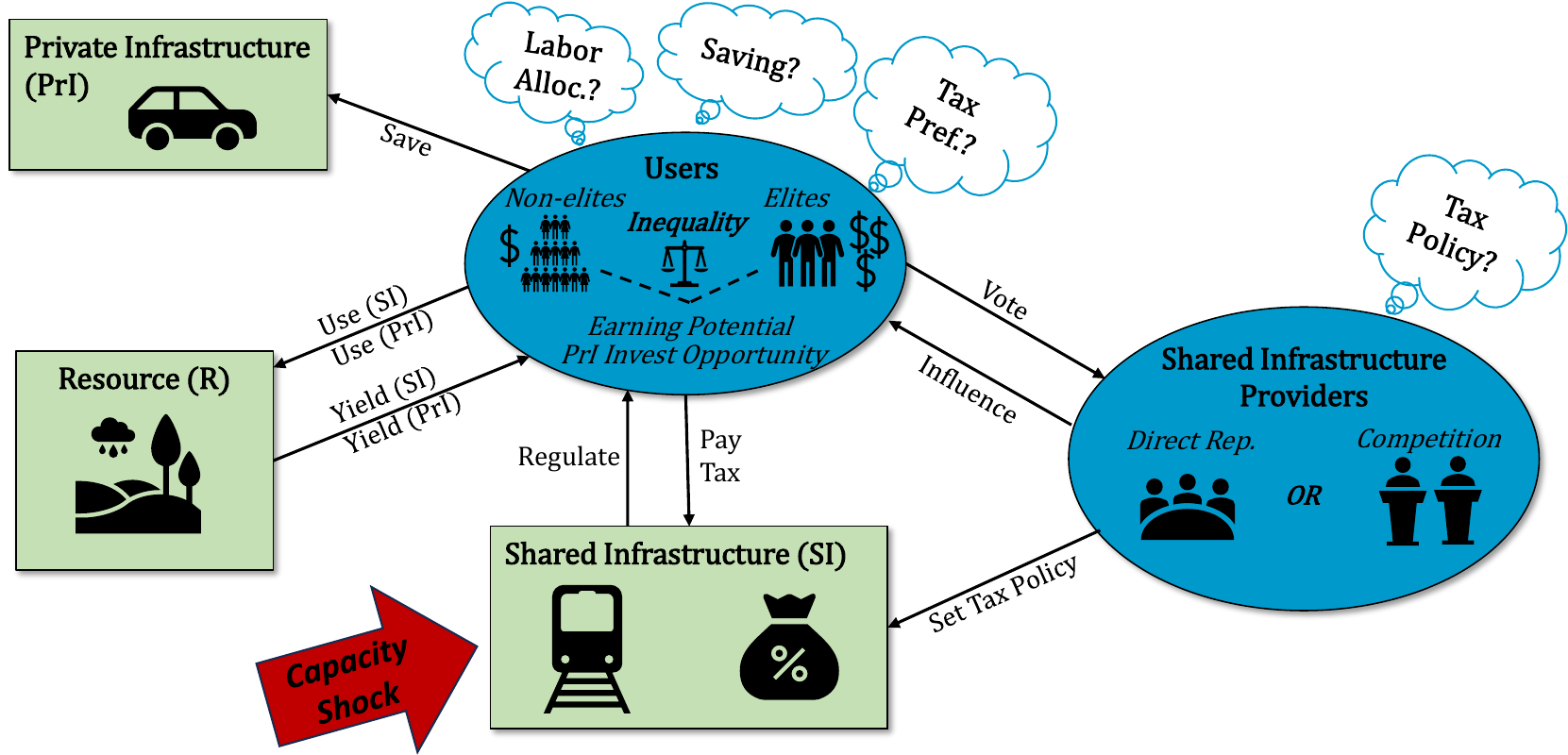}
    \caption{Summary of model components using the layout defined by the Coupled Infrastructure Systems Framework \cite{Muneepeerakul2017,Muneepeerakul2020}. Arrows indicate flows of material and information and thought bubbles indicate dynamic decisions made by users and shared infrastructure providers.}
    \label{fig:ModelOverview}
\end{figure*}

We build on a well-studied continuous dynamical system application of the Coupled Infrastructure Systems (CIS) framework \cite{anderies_framework_2022} (referred to as \emph{MA model}) that captures the effect of external wage opportunities on the willingness of users to participate in a shared infrastructure system to process a dynamic resource into yields \cite{Muneepeerakul2017,Muneepeerakul2020}. We proceed through model development and analysis in iterative steps to ensure that the implications of each added mechanism are carefully considered before adding complexity \cite{siegel_analyzing_2018}, finally reaching the overall model structure depicted in Figure \ref{fig:ModelOverview}. Shared infrastructure is a stock variable, $I^s$, whose capacity increases with investment and decreases with depreciation. We assume a static, exogenously determined resource ($\dot{R}=0$) to focus on the political and infrastructure dynamics. Over time, users update their proportional allocation of labor into the shared system ($l \in [0,1]$) based on the perceived yield of shared infrastructure use per unit of labor allocated.

Shared infrastructure investment is funded through revenues collected as a proportion of user-generated yields, or the tax rate, $\tau$. The key political-economic dynamic of the model is updating $\tau$ by the shared infrastructure providers. This can also be thought of as the intended level of government spending because tax policy accommodates intended spending \cite{persson_political_2000}. We model the tax-setting process as either a direct aggregation of voter preferences (\emph{DirectAgg} models) or an indirect mediation of voter preferences through election of competiting candidates (\emph{PolComp} models). 

\begin{figure*}
    \centering
    \includegraphics[width=\linewidth,trim={0 0 0 0},clip]{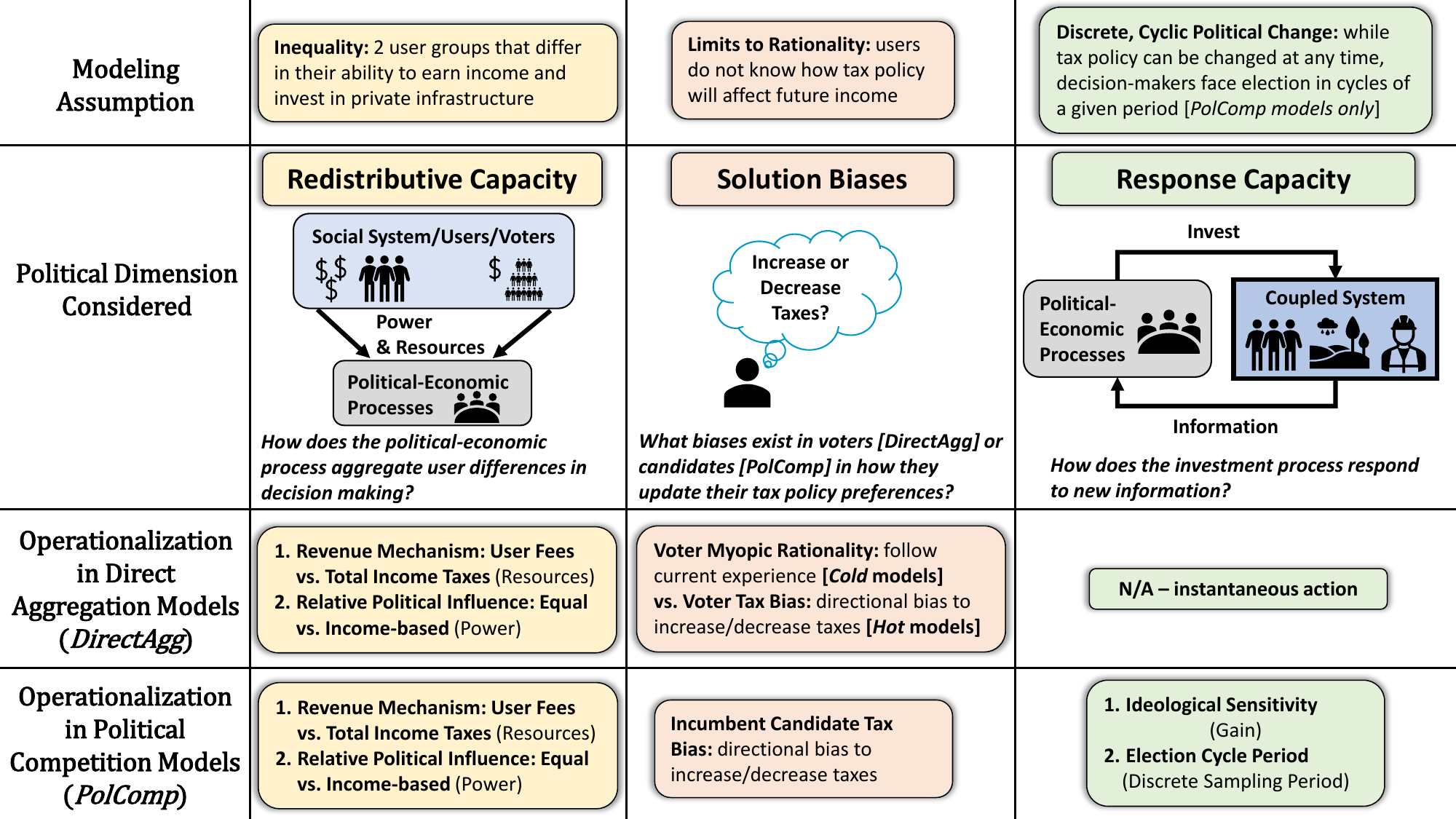}
    \caption{Political dimensions considered in model development and analysis. We begin (first row)  with three central assumptions that add complexity to the infrastructure system: (i) inequality, (ii) limits to rationality, and (iii) discrete and cyclic political change. We then (second row) consider the structural features of the model that implement the assumptions within the model's dynamics and characterize their effect on infrastructure politics: (i) redistributive capacity, (ii) solution biases, and (iii) response capacity. In the last two rows, we explain how we operationalized each political dimension in the \emph{DirectAgg} and the \emph{PolComp} models, respectively.}
    \label{fig:typetree}
\end{figure*}

In a \emph{DirectAgg} model, there are no intermediary agents (e.g., politicians) between voter preferences and policy decisions, just an aggregation rule. We consider two rules (Figure \ref{fig:typetree}): median voter, which assumes equal voter influence \cite{downs_economic_1957,persson_political_2000} (\emph{DirectAgg-MV model}), or elite capture \cite{acemoglu_theory_2001,acemoglu_why_2000} (\emph{DirectAgg-EC model}), where users with higher income set policy. For idealized historical examples, one might consider Tocqueville's description of early American assemblies \cite{tocqueville_democracy_1838} for \emph{DirectAgg-MV} or British ``oligarchy" before extending franchise in 1832 \cite{acemoglu_why_2000} for \emph{DirectAgg-EC}. Voter preferences, under \emph{DirectAgg}, update according to a ``cold" calculation of immediate yields (will a marginal increase in the tax rate increase or decrease yields) or a ``hot" calculation based on pre-existing bias (that is, whether they believe tax decreases or increases are always desirable) when their yields fall below prior expectations, referred to as \emph{error} (Figure \ref{fig:typetree}). We derive the ``cold" and ``hot" learning algorithms from prior models of retrospective voting \cite{duggan_political_2017,healy_retrospective_2013,persson_political_2000} and motivated reasoning \cite{little_motivated_2022,druckman_evidence_2019,redlawsk_hot_2002}, respectively.

In \emph{PolComp} models, we define two candidates who, like ``hot" learning users, have opposing views on the direction taxes should be changed (Figure \ref{fig:typetree}). Our approach, based on computational party competition modeling \cite{de_marchi_voter_2003,Kollman2006}, assumes that candidates are motivated by both ideological preference and winning election \cite{wittman_candidate_1983}, but we adopt a directional understanding of ideology (e.g., tax increases versus decreases) as opposed to ideal points (e.g., a certain desired level of taxing), which are difficult to measure and limit generalizability. We assume that candidates do not know voter preferences \cite{kollman_adaptive_1992,fowler_dynamic_2005}, so election results are the primary signal of voter preferences. Candidate platform updating involves a combination of general attraction (weighted by $\sigma_A$) towards status quo policy, but also a repulsion (weighted by $\sigma_R$) from status quo policy in their biased direction, when there is error. 

Under \emph{PolComp}, voters update their preferences in the direction of candidate platforms as a function of (i) the distance of the platform from their current preference \cite{kollman_adaptive_1992,axelrod_preventing_2021}, and (ii) the magnitude of current error, where incumbents are punished but challengers are rewarded \cite{duggan_political_2017,ingberman_reputational_1989,healy_retrospective_2013}. Every $T_e$ time steps, an election occurs where the candidate with the most votes - determined by the distance of their platform to voter preferences - gains control of tax-setting and enacts their platform. Like \emph{DirectAgg}, we consider two distributions of political influence (Figure \ref{fig:typetree}): equal (\emph{PolComp-Eq}) and income-based (\emph{PolComp-Inc}). \emph{PolComp-Eq} reflects the ``one person, one vote" distribution, often invoked as the ideal in democratic republics. However, with the rise of campaign spending and organized corporate lobbying, \emph{PolComp-Inc} reflects a ``one dollar, one vote" distribution noted in OECD countries with high inequality \cite{karabarbounis_one_2011}. Note, in our continuous model, platforms update in-between elections, so once in office, an incumbent can update their policy from the elected platform based on changing system state. 

If one were to conceptualize tax setting as a controller, $\sigma_R$ and $\sigma_A$ serve as the \emph{gains} in translating perceived errors into policy while $T_e$ serves as the \emph{discrete sampling period} for political change. Together, they define the \emph{response capacity} of the political-economic system (Figure \ref{fig:typetree}). 

In addition to labor allocation, users make a dynamic decision to save a proportion $s$, of their post-tax income and invest in private infrastructure $I^p$, which follows the same depreciation and investment dynamics as $I^s$. This decision is similar to labor allocation in that it is based on whether a perceived marginal increase in $s$ will increase or decrease yield. We divide users into two groups (indexed by $g$) that differ according to three parameters: (i) their relative share of the user population, $n_g$, (ii) their ability to invest in private infrastructure, $\mu^p_g$, and (iii) their income potential, $\phi_g$.

Lastly, we consider an additional feature of redistributive capacity: whether the use of private infrastructure is taxed (Figure \ref{fig:typetree}). For each political-economic model, we compare the effects of user fee or total income taxing models. User fees only tax shared infrastructure use while total income taxing taxes both shared and private infrastructure use. 

\begin{table*}[!t]
  \centering
     \caption{Key Parameters and Summary Variables. Unit symbols:  $T$: time,  $N$:  population,  $R$: resource level, $I$: infrastructure capacity, \$: welfare, N/A: none.}
    \begin{tabular}{clcc}
    \toprule
       Symbol  & Name & Units & Default \\
       \midrule
      $n_g$  & Number of users in group $g$ & $N$ & $[200,800]$ \\
       $\phi_g$ & Earning potential of group $g$ & $\frac{\$}{R}$ & $[30, 7.5]$ \\
       $\mu$ & Share infrastructure investment effectiveness & $\frac{I}{\$}$ & $0.001$ \\
       $\mu^p_g$ & Private infrastructure investment effectiveness for group $g$ & $\frac{I}{\$}$ & $[0.0015, 0.0005]$ \\
       $\delta$ & Infrastructure decay rate & $\frac{1}{T}$ & $0.1$ \\
       $h$ & Maximum harvesting rate & $\frac{1}{NT}$ & $0.0025$ \\
       $w$ & Subsistence income without infrastructure & $\frac{\$}{NT}$ & 0.1 \\
       $\psi$ & Is private use taxed? & N/A & $0$ \\
       $\alpha$ & Relative elite political influence & N/A & $0$ \\
       $\xi$ & Habituation rate of users & N/A & 0.1 \\
       $\beta^l$ & Labor allocation sensitivity & $\frac{N}{\$}$ & 0.15 \\
       $\beta^s$ & Savings rate sensitivity & $\frac{N}{\$}$ & 0.015 \\
       $\beta^{\tau}_1$ & Cold cognition sensitivity & $\frac{N}{\$}$ & 0.06 \\
       $\beta^{\tau}_2$ & Hot cognition sensitivity & $\frac{N}{\$}$ & 0.06 \\
       $\theta_g$ & Ideological bias for group $g$ & N/A & $[-1,1]$ \\
       $\check{\theta}_q$ & Ideological bias for candidate $q$ & N/A & $[-1,1]$ \\
       $\sigma^A$ & Weight of status quo attraction in platform updating & $\frac{N}{\$}$ & 0.1 \\
       $\sigma^R$ & Weight of ideological repulsion in platform updating & $\frac{N}{\$}$ & 0.15 \\
       $T_e$ & Time between elections & $T$ & 4 \\
    \bottomrule 
    \end{tabular}
\label{tab:params_dimensionless}
\end{table*}

We emphasize that our approach can be applied to many different systems, provided that the system's embedding space is specified. By embedding space, we refer to all infrastructure treated as external to the modeled dynamics. For example, users might consider the choice to take public transport ($I^s$) or to use their car ($I^p$), which may take a mixed strategy ($0<l<1$). The embedding space would be all other shared and private infrastructure (e.g., roads, communication networks, rules, etc.) relevant to their choice. 

\subsection{Measuring Robustness}

We conducted deterministic and stochastic sensitivity analyzes to measure the robustness of the shared system. The deterministic analysis considered a step-change shock in two dimensions: (i) a decrease in shared infrastructure capacity, $\Delta^{I_s}$, and (ii) an increase in elite opportunity to invest in private infrastructure, $\Delta^{\mu^p_1}$. $\Delta^{I_s}$ may be interpreted as biophysical (e.g., disruptions from natural disasters) or socioeconomic (e.g., cyber attacks or economic downturns). We report $\Delta^{I_s}$ as the magnitude of the proportional change in the capacity of shared infrastructure (all in the negative direction). Common examples of $\Delta^{\mu^p_1}$ include the introduction of new technology (e.g., new tube well infrastructure in a canal system) and biophysical (e.g., increased storms encourages rainwater harvesting) or socioeconomic (e.g., new markets) changes. We report it as a proportional change in magnitude normalized to $\mu$.

The stochastic analysis considered a shock regime for $\Delta^{I_s}$ drawn from a Poisson process with a given mean shock period, $T_s$, and mean magnitude, $a$ \cite{krueger_resilience_2019}. We sample 400 shock time series from a shock regime ($T_s$ and $a$ combination) when performing the sensitivity analysis. In the stochastic analysis, because the initial incumbent is not necessarily in power when a shock occurs, we randomly set the initial incumbent at the start of each model run with simple random sampling. 
  
The main outcome variable of interest for assessing robustness is $I^s$, measured at the end of a 400 time-step run. To calculate robustness, we adopt the safe operating space approach \cite{anderies_knowledge_2019}, where we note the proportion of runs, under a given political configuration (Figure \ref{fig:typetree}), where the shared infrastructure is preserved ($I^s>0$). A political configuration is more robust if it has a larger safe operating space. 

\section{Results}
In our robustness analysis of various political contexts (Figure \ref{fig:typetree}), we highlight how the presence of inequality, information constraints, and political competition requires consideration of redistributive mechanisms, biases, and response capacity within the political-economic context. The key outcome of interest is the persistence of shared infrastructure ($I^s>0$) at the end of a model run.

\subsection{Four Possible Post-Shock States}

\begin{figure}
    \centering
    \includegraphics[width=\linewidth]{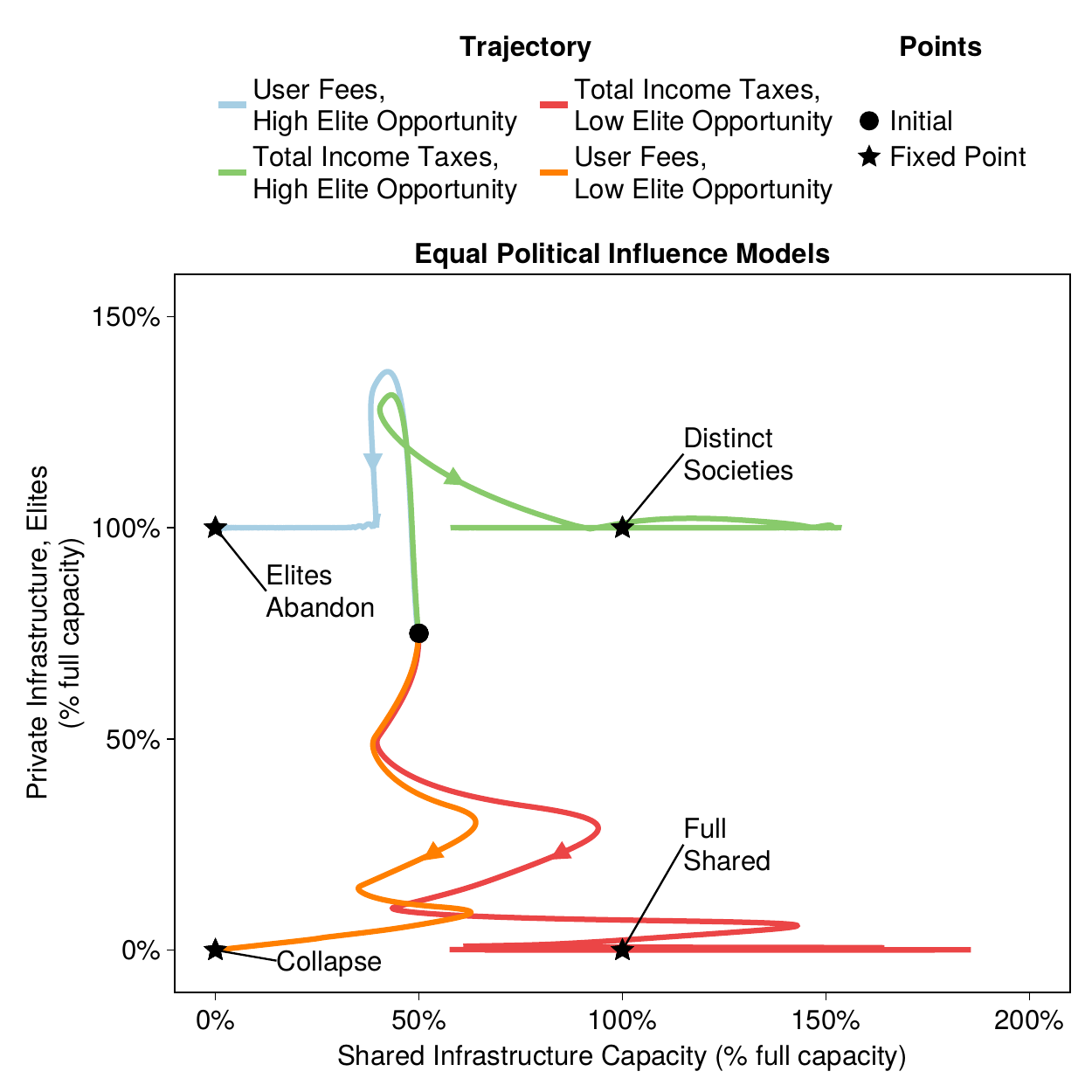}
    \caption{There are four post-shock states for the system: Full Shared (full shared infrastructure capacity, no private capacity), Elites Abandon (no shared capacity, full private capacity for elites), Distinct Societies (full shared capacity, full private capacity for elites), and Collapse (no infrastructure capacity). We plot four example trajectories of the post-shock states from the same initial shock of a 50\% reduction in shared infrastructure capacity that differ by whether there is a user fee or total income tax in place and the private infrastructure investment opportunity for elites. The trajectories that lead to Distinct Societies and Full Shared states provide examples of the limit cycle behavior created by overshoots in the competitive electoral cycle.}
    \label{fig:phase}
\end{figure}

In response to a shock, the system can move to one of four post-shock states, the four fixed points in the dynamical system (Figure \ref{fig:phase}): Full Shared, Collapse, Elites Abandon, and Distinct Societies. In Full Shared, all users fully participate in the shared system ($l_1=l_2=1$), and elites do not have private infrastructure ($I^p_1=0$). Although this fixed point exists at full shared capacity ($I^s = \bar{I}$), there is the potential for limit cycles to occur where $I^s$ oscillates in partial and full capacities. This is due to oscillations between high and low taxation created by candidates overshooting in their platform updating, especially when there are large shocks and high ideological sensitivity ($\sigma_R$). However, we also note that such swings in tax policy are a feature of US history \cite{baker_tax_2020} and the international cyclical behavior of tax rates \cite{vegh_how_2015,franzese_jr_electoral_2002}.

In Collapse, both shared infrastructure and private infrastructure collapse ($I_s=I^p_1=0$), and all users revert to subsistence income ($w$). In Elites Abandon, elites fully exit the shared system ($l_1=0$) and develop their private infrastructure to full capacity ($I^p_1 = \bar{I}\tilde{n}_1$), which collapses the shared system ($I^s=0$) and drives non-elites to subsistence. We do not see limit cycles because the private savings dynamics converge smoothly (Equation \ref{eq:savings}). In Distinct Societies, both elite private infrastructure and shared infrastructure co-exist at full capacity ($\frac{I^p_1}{\tilde{n}_1}=I^s = \bar{I}$) with elites fully using private infrastructure ($l_1=0$) and non-elites fully using shared infrastructure ($l_2=1$). 

\subsection{Inequality \& Redistribution}

We consider inequality through the opportunity to invest in private infrastructure, $\mu^p_g$, or the ability to convert well-being units (\$) into private infrastructure. Without inequality, trivial cases occur where the robustness of shared infrastructure depends on the relative effectiveness of investment in shared versus private infrastructure (i.e., robustness requires $\mu^p_g < \mu$). We initialize non-elite opportunity ($\mu^p_2$) to be less than that of shared infrastructure (Table \ref{tab:params_dimensionless}), but elite opportunity ($\mu^p_1$) is a shock of interest in the deterministic analysis. 

By inspection of the labor allocation replicator equation (Equation \ref{eq:labor}), increasing $\mu^p_1$ increases private infrastructure capacity ($I^p_1$), and if the shared infrastructure shock ($\Delta^{I_s}$) is high enough, private infrastructure will appear more lucrative to elites, who will then exit the shared system. Under user fees, an Elites Abandon outcome will occur if the user fee is not raised high enough to compensate for lost revenue. Thus, our first proposition is as follows, 

\begin{quote}
    \textbf{Proposition 1:} Increasing private infrastructure investment inequality decreases the robustness of a shared infrastructure system to capacity shocks under user fee revenue generation.
\end{quote}

\begin{figure*}[!t]
    \centering
    \includegraphics[width=0.9\linewidth]{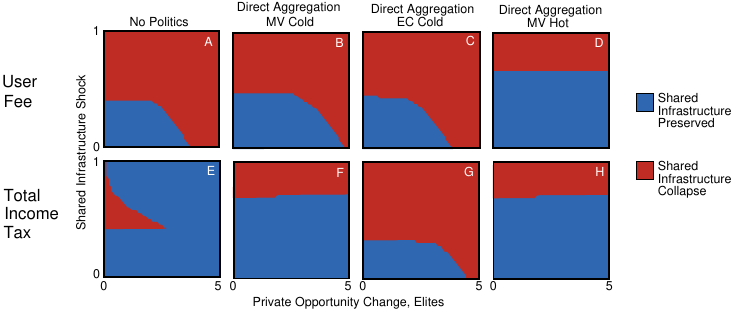}
    \caption{The \emph{NoPolitics} and \emph{DirectAgg} models differ in their safe operating spaces under user fee (A-D) and total income taxing mechanisms (E-H). Each panel contains the robustness results (whether shared infrastructure was preserved) for a given political model (\ref{fig:typetree}) over the two-dimensional shock space (shared infrastructure capacity and elite private opportunity). We do not plot results from the hot cognition and elite control (\emph{EC-Hot}) model because the shared infrastructure system always collapses. Cold cognition, user fee models have a smaller region of shared infrastructure preservation (lower robustness) at high elite opportunity (B-C). Total income taxing prevents this effect when there is median voter influence (F), but under elite control (G), elites remove the tax when they abandon the shared system. Under hot cognition and median voter influence (D, H), system recovery is only limited by available taxable income, which is compromised at high shared infrastructure shock levels.}
    \label{fig:SOS_pol_nocomp}
\end{figure*}

This proposition holds for the \emph{NoPolitics} and cold cognition \emph{DirectAgg} models (Figure \ref{fig:SOS_pol_nocomp}a-d). Because the initial tax rate is set at a maintenance point, an increase in the tax rate is required to recover shared infrastructure from $\Delta I^s$. Under \emph{DirectAgg-MV-Hot} (Figure \ref{fig:SOS_pol_nocomp}d), the shared system bounces back regardless of $\Delta\mu^p_1$ because non-elites are biased to seek tax increases, but if $\Delta I^s$ is high enough, there is not enough income available to invest in shared infrastructure. We do not plot the results for \emph{DirectAgg-EC-Hot} because the tax decreasing bias causes the system to always collapse. Under cold cognition (Figure \ref{fig:SOS_pol_nocomp}b-c), high $\Delta \mu^p_1$ tempts elites to exit, and non-elites are hesitant to increase tax levels because they perceive a negative effect on short-term consumption. 

User fees are commonly promoted for the efficiency benefits of price signals \cite{pagano_funding_2011,alm_financing_2015}, but Proposition 1 points to a performance-fragility trade-off \cite{Anderies2013a} created by elite exiting. This is where redistributive revenue mechanisms can help. Total income taxing can increase the robustness of the system because elites are still taxed when they opt for private infrastructure (Figure \ref{fig:SOS_pol_nocomp}e-h). A common example is the use of property taxes to fund public schools, regardless of whether a household opts for private schools. However, we note that the presence of elite capture (\emph{DirectAgg-EC-Cold}, Figure \ref{fig:SOS_pol_nocomp}g) voids this effect because elites eliminate the tax after exiting. Thus, we provide the following proposition: 

\begin{quote}
    \textbf{Proposition 2:} Total income taxing increases the robustness of shared infrastructure systems to private infrastructure investment inequality, provided that the system is insulated from elite capture.
\end{quote}

\begin{figure*}[!t]
    \centering
    \includegraphics[width=\linewidth]{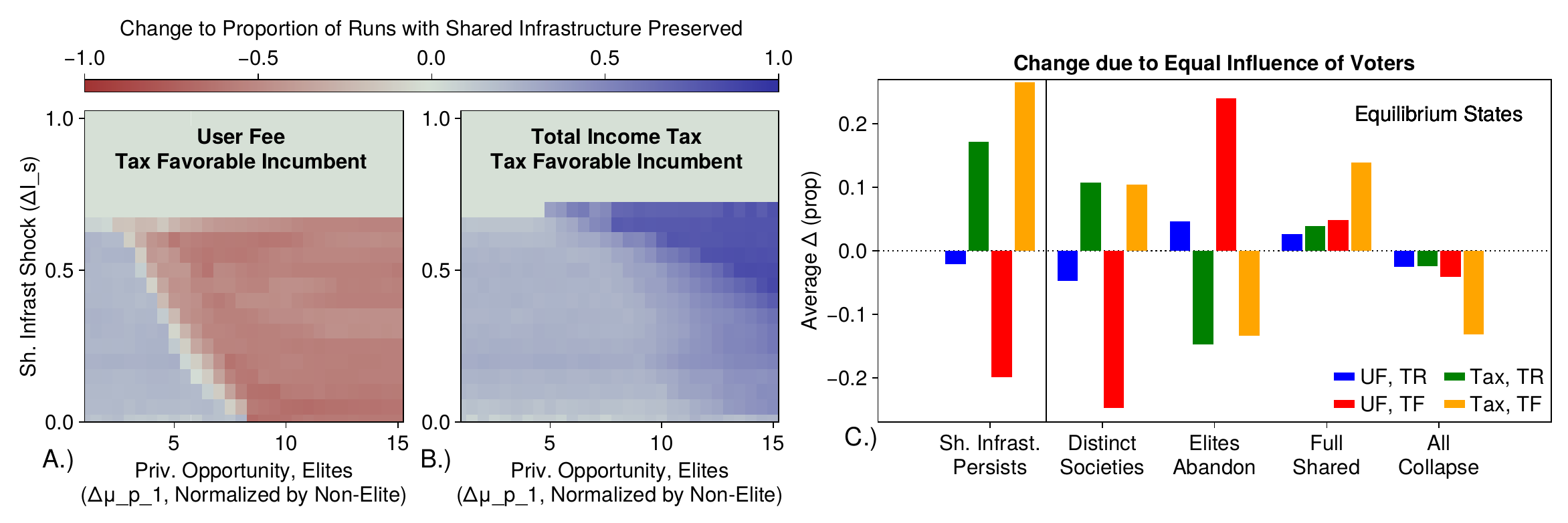}
    \caption{The two dimensions of redistributive capacity, resources and power (Figure \ref{fig:typetree}), have an interactive effect on the robustness of shared infrastructure to capacity and elite opportunity shocks. We present the results from the four-dimensional sensitivity analysis ($T_e$, $\sigma_R$, $\Delta^{I_s}$, $\Delta^{\mu^p_1}$) conducted on each political model (\ref{fig:typetree}) as a projection onto the shock space ($\Delta^{I_s}$ vs. $\Delta^{\mu^p_1}$) in A \& B, showing the average change to robustness by having equal influence for each shock combination. In C, we plot the total average change over the entire sensitivity analysis to robustness, as well as the relative frequency of each equilibrium state due to having equal voter influence. When under a user fee model (A), equal influence decreases robustness when there is high elite opportunity ($\Delta^{\mu^p_1}$). However, total income taxing (B) benefits from equal political influence. This effect occurs for both initial incumbents (C) but is less strong for a tax-repulsed initial incumbent.}
    \label{fig:redistrib}
\end{figure*}

Propositions 1 and 2 also apply to the \emph{PolComp} models. In the deterministic analysis, the \emph{PolComp} models illustrate a critical distinction between the redistribution of resources (revenue) and influence (power). In line with Proposition 2, the robustness of a shared system with total income taxation is greater when voters have equal political influence (\emph{PolComp-Eq}) than when political influence is based on income (\emph{PolComp-Inc}) (Figure \ref{fig:redistrib}b). This is because in the \emph{PolComp-Inc} model, like \emph{DirAgg-EC-Cold}, when elites leave, they vote for candidates that decrease taxes, collapsing shared infrastructure and creating more Elites Abandon and less Discrete Societies outcomes (Figure \ref{fig:redistrib}c). 

However, the opposite effect occurs with user fees (Figure \ref{fig:redistrib}a, c). Because non-elites have less available income, they are more hesitant to vote for tax-favoring candidates when elites exit and increase the revenue burden. Under \emph{PolComp-Inc}, because exiting elites are not affected by user fees, they are ambivalent and do not change it, leading to more Distinct Societies outcomes (Figure \ref{fig:redistrib}c).

In the stochastic analysis, we ensure that there is potential for elite exiting by maintaining a cap on $l_1$ ($l_1 \leq 0.9$) if shocks occur later in the model run (see Appendix for justification).

\subsection{Information Constraints \& Biases}

We assume that users and candidates do not know how tax policy will affect future income. The most generous voter knowledge model used in our analysis is \emph{DirectAgg-Cold}, where voters follow their immediate experience of how tax policy affects income in a replicator dynamic similar to the labor allocation and savings rate decisions. In the \emph{DirectAgg-Hot} and \emph{PolComp} models, voters or candidates, respectively, follow their directional bias when updating their tax preference. 

Assumptions about the learning process play a critical role in the robustness of the shared infrastructure system. This is clear in the \emph{DirectAgg-Hot} models, where tax-favoring biases are required to recover the needed revenue (Figure \ref{fig:SOS_pol_nocomp}d,h). Similarly, in the \emph{PolComp} models, candidate biases intuitively aid or resist tax increases (see Appendix for the effect of the initial incumbent's bias on robustness).

In \emph{DirectAgg-Cold} models, if non-elites have more political influence, a tax rate of $\tau=0.2$ would be required to maintain shared infrastructure when elites exit within a user fee system. If the voters operated under an infinite time horizon, the needed tax increase would appear favorable, but under cold cognition, non-elites are hesitant to increase taxes to that level because the short-term threat to consumption is too high (Figure \ref{fig:SOS_pol_nocomp}b). Thus, we offer the following proposition:

\begin{quote}
    \textbf{Proposition 3:} The effect of increasing representation (i.e., redistribution of political influence) on the robustness of shared infrastructure to capacity shocks depends on (i) the direction of voter or incumbent candidate biases and (ii) the time horizon in preference updating.  
\end{quote}

\subsection{Political Competition\& Response Capacity} 

In the \emph{PolComp} models, because candidates serve as imperfect intermediaries between voter preferences and policy, we evaluate how the response capacity in the political economic system (Figure \ref{fig:typetree}) affects the robustness of the shared infrastructure. While the incumbent politician can change the tax at any point (real-time policy response), elections create discontinuities in the \emph{political} response of the system (changing the incumbent).  

\begin{figure}[ht]
    \centering
    \includegraphics[width=\linewidth]{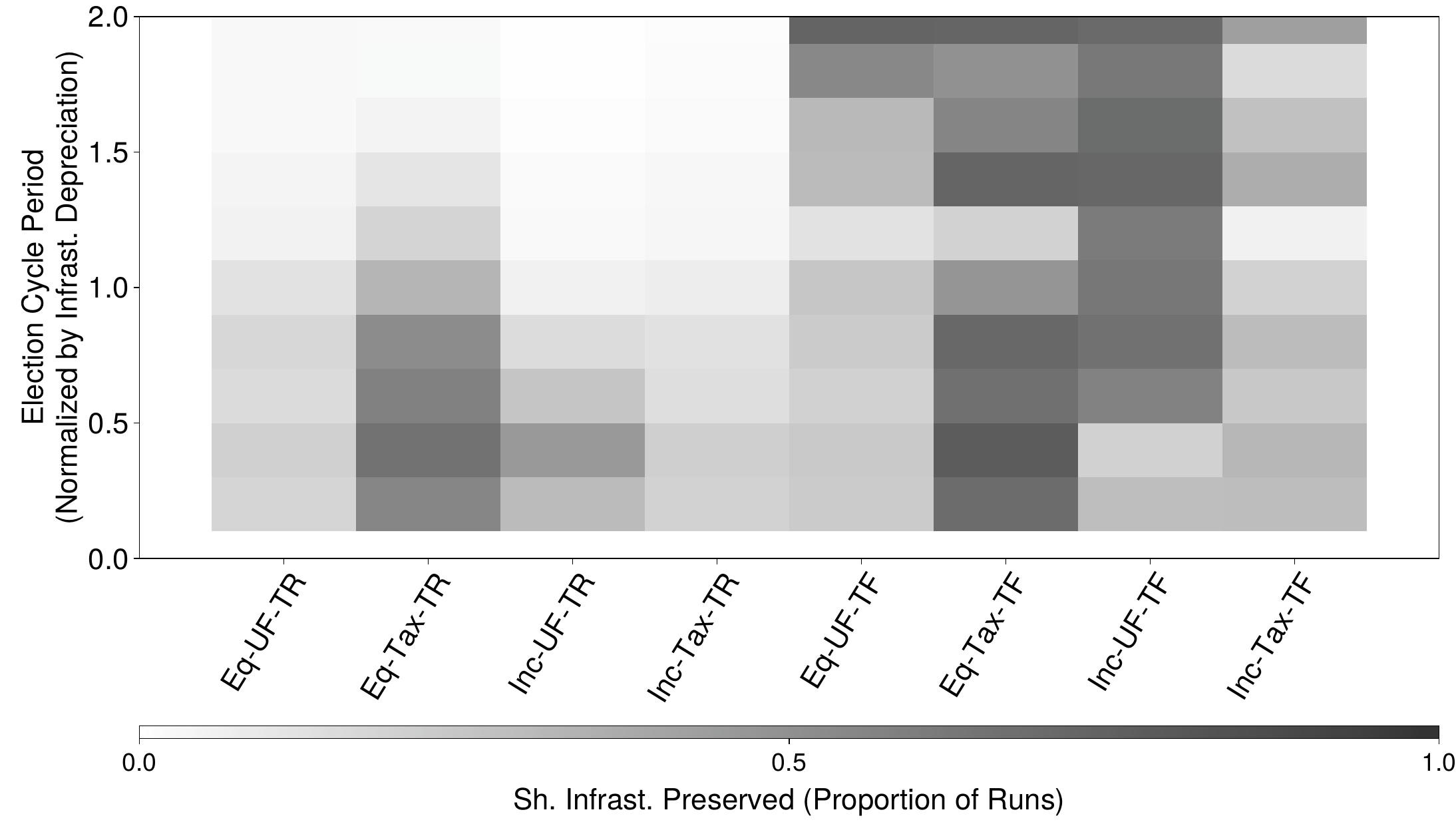}
    \caption{Longer election cycle periods benefit robustness when the initial incumbent is tax-favoring. We present the results from the four-dimensional sensitivity analysis ($T_e$, $\sigma_R$, $\Delta^{I_s}$, $\Delta^{\mu^p_1}$) as a projection onto the $T_e$ dimension. Each column represents a \emph{PolComp} model distinguished by equal (Eq) versus income-based (Inc) influence, user fees (UF) versus total income taxing (Tax), and tax-repulsed (TR) versus tax-favorable (TF) initial incumbents. Shade darkness indicates the proportion of runs where the shared infrastructure system persisted. TF incumbents benefit from higher election cycle periods to give their policies time to develop while a TR incumbent needs to be removed quickly before recovery can begin.}
    \label{fig:response_heat}
\end{figure}

In the deterministic analysis, the effect of $T_e$ and $\sigma_R$ on robustness is mediated by the initial incumbent (Figure \ref{fig:response_heat}). This is because a system with a tax-repulsed incumbent benefits from a quick response (low $T_e$ and high $\sigma_R$) to get the incumbent out of office, while a system with a tax-favoring incumbent benefits from a slower response (high $T_e$ and low $\sigma_R$) to recover the system without generating backlash from acting too quickly. Thus, to provide general insight on response capacity, we use the stochastic analysis to control for the incumbent in office at the time a shock occurs and randomize the initial incumbent for each model run.

\begin{figure}[ht]
    \centering \includegraphics[width=0.75\linewidth]{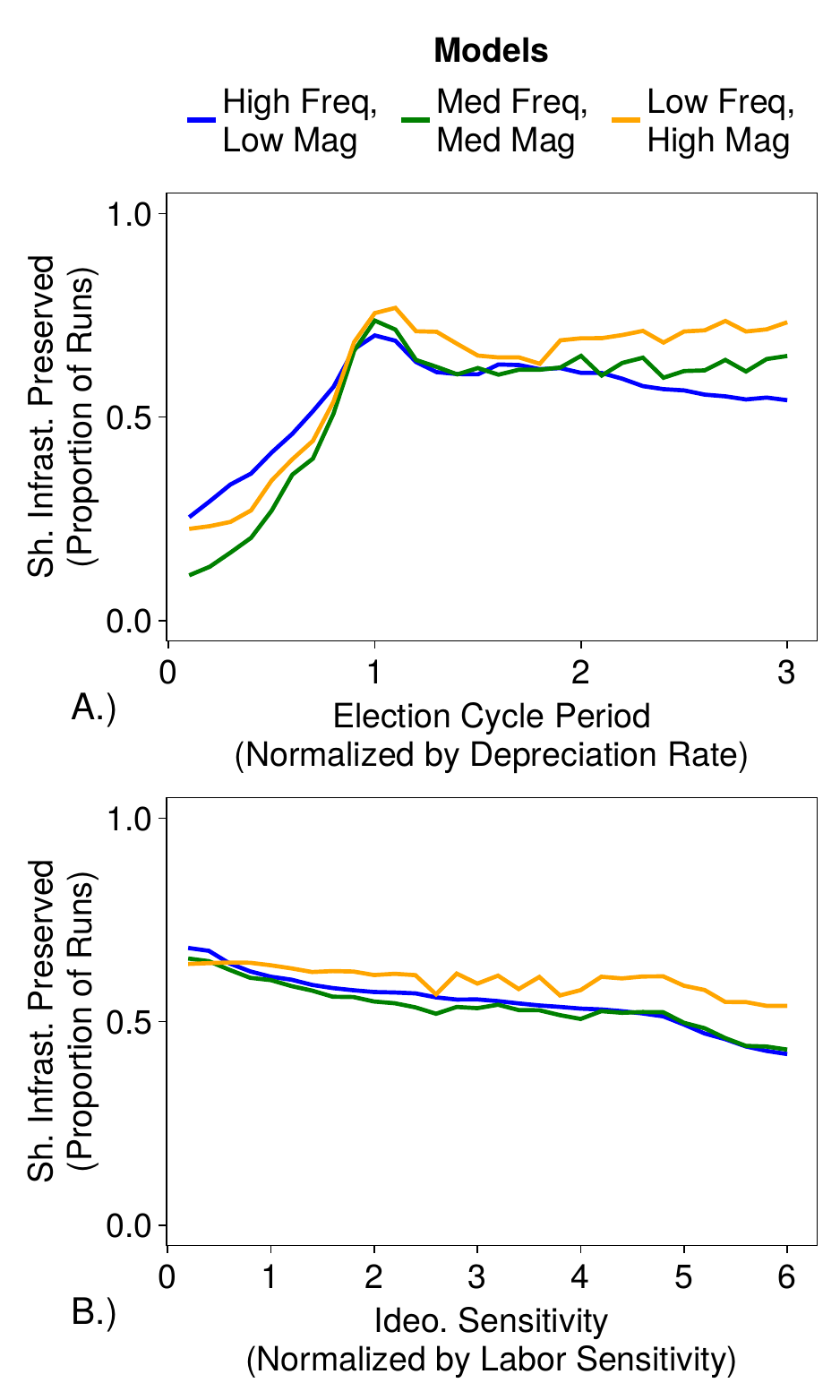}
    \caption{ The robustness of a shared infrastructure system where tax policy is subjected to political competition is dependent on the election cycle period, $T_e$, and the ideological sensitivity of the candidates, $\sigma_R$. We plot results from the stochastic analysis of $T_e$ and $\sigma_R$, reporting the average proportion of runs where the shared infrastructure system was preserved for a given value of $T_e$ (A) and $\sigma_R$ (B) normalized to the depreciation rate and labor sensitivity, respectively. To limit the computational load of the analysis, we conducted the sensitivity analysis on three representative shock regimes of high frequency-low magnitude ($T_s = 8$, $a=0.1$), medium frequency-medium magnitude ($T_s=50$, $a=0.25$), and high frequency-low magnitude ($T_s=150$, $a=0.5$). There is a nonlinear relationship between $T_e$ and robustness where below a normalized $T_e$ of 1, increasing $T_e$ greatly benefits robustness and above 1, increasing $T_e$ tends to decrease robustness at a lower rate under the high frequency, low magnitude shock regime (A). For the other two shock regimes, robustness appears to level off beyond 1 (A). Increasing $\sigma_R$, across the range explored, tends to decrease robustness on average, but the effect for the low frequency, high magnitude shock regime is less significant (B).}
    \label{fig:MS_response}
\end{figure}

From the stochastic analysis, we plot the robustness of the shared system for a given $T_e$ setting, averaged across all $\sigma_R$ settings explored, for three representative shock regimes (Figure \ref{fig:MS_response}). When $T_e$ is less than $\frac{1}{\delta}$, there are significant benefits to increasing $T_e$, but as $T_e$ increases beyond this level, there is a gradual decrease in robustness. At low $T_e$ values, incumbents do not have enough time to impact the system, regardless of their bias, because the electoral process reacts prematurely. At high $T_e$ values, the electoral process has to wait too long to react to undesirable outcomes, particularly when shock magnitudes are lower and their frequency is higher (blue line in Figure \ref{fig:MS_response}a). This is because a more responsive electoral system addresses low-magnitude shocks before they can accumulate. When a system experiences acute shocks (low frequency, high magnitude), the effect is sudden, so the system's fate is dependent on the incumbent's biases (Proposition 3), rather than electoral responsiveness. We caution that the $T_e$ that maximizes robustness is a function of the infrastructure system, which is why we normalize it to the depreciation rate. Thus, we offer the following proposition for the effect of $T_e$: 

\begin{quote}
    \textbf{Proposition 4:} Increasing the election cycle period for political control of tax policy increases the robustness of a shared infrastructure system to capacity shocks up to a point, beyond which, robustness decreases with increasing election cycle period, particularly when the system experiences more frequent, lower magnitude shock regimes. 
\end{quote}

On the other hand, when we plot the effect of $\sigma_R$, averages over all $T_e$ settings (Figure \ref{fig:MS_response}b), there is a negative effect of increasing $\sigma_R$ on robustness, due to the potential for political overshoot and electoral backlash when candidates are too ideologically sensitive. Thus, we offer the following proposition regarding $\sigma_R$:

\begin{quote}
    \textbf{Proposition 5:} Increasing the ideological sensitivity of candidates competing for political control of tax policy decreases the robustness of a shared infrastructure system to capacity shocks.
  \end{quote}

\section{Discussion \& Conclusion}

We present a dynamical systems approach to understand the role of politics (Figure \ref{fig:typetree}) in the robustness of shared infrastructure systems when there are opportunities for private infrastructure. The analysis yields multiple propositions that formalize the interactive effects between inequality in resources and influence, revenue-generating mechanisms, assumptions about the learning process, political competition, and the robustness of a shared infrastructure system. 

To prepare our critical infrastructure systems for the uncertain social and environmental futures of the Anthropopocene, our focus must move beyond the design of infrastructure systems themselves and account for the institutions that guide the political process of infrastructure investment \cite{Anderies2013b,deslatte_assessing_2022}. In addition to increasing the frequency and magnitude of shocks \cite{chester_infrastructure_2020}, ongoing social and technological progress has introduced new possibilities for private infrastructure that may not be viable for all users. Examples include electric vehicles \cite{romero-lankao_inequality_2022}, private water infrastructure \cite{deitz_plumbing_2019}, private solar energy generation and storage \cite{sovacool_towards_2022}, private schooling and remote instruction \cite{reynolds_digital_2022}, and cyberinfrastructure \cite{zheng_inequality_2021}. 

Our findings (Propositions 1-2) emphasize that inequality in private infrastructure opportunities can compromise a shared infrastructure system that the rest of users rely upon unless mechanisms exist to redistribute lost revenue (e.g., taxing private use) \emph{and} ensure equal political influence. One mechanism without the other threatens the shared system (Figure \ref{fig:redistrib}). In line with Acemoglu and Johnson's historical analysis \cite{acemoglu_power_2023}, the effect of a new technology on collective well-being is not only a product of the technology itself, but of the institutional landscape the technology enters as inequalities in access and power are inevitable. This is precisely the motivator for \emph{shared} infrastructure, using economies of scale to provide essential capacity to diverse populations \cite{frischmann_infrastructure_2012,levin_governance_2022}.

Our findings also highlight the complex effect of electoral competition on shared infrastructure systems. Political responses through the electoral process are inherently imperfect due to discrete sampling that rarely coincides with the frequency of shocks or the time needed to recover a damaged system. In fact, a former national security advisor lamented,

\begin{quote}
   the benefits that flow to the American people [from infrastructure policies] will be measured in decades, and political cycles are measured in years \cite{klein_opinion_2025}.
\end{quote}

To better illustrate this challenge, suppose that a home air conditioning system can only be changed to heating or cooling modes at certain times of the day. If an impromptu party begins in-between the designated times, flooding the house with heat-emitting people, the comfort level will depend on whether the system was previously set to heat or cool. In electoral systems, the designated times to change the system are elections. Furthermore, to navigate the complexity of government spending, political leaders often revert to directional biases (heat or cool the system), such as the debate to increase investment in public transit or privatize it. Faced with this challenge, one might argue to maximize the frequency of elections to maximize responsiveness. However, with infrastructure, if elections are more frequent than investment implementation, there is strong potential for premature electoral backlash (quote above) \cite{mullin_local_2022,glaeser_political_2018}. 

A dynamical systems model is valuable because the issue of interest is the nonlinear interaction of infrastructures and multiple political-economic decisions made at different relative rates. We contribute to computational modeling of party competition \cite{de_marchi_voter_2003,Kollman2006} and polarization \cite{axelrod_preventing_2021,levin_polarization_2024} by coupling existing approaches to a dynamic infrastructure system that provides stakes for political outcomes. This also yields two key findings pertaining to political response capacity. First, we illustrate the parabolic relationship between the election cycle period and robustness, where the period must be set long enough for policy implementation, but not so long as to delay needed political feedback (Proposition 4). This tuning question is very context-driven, and multiple policy questions in real systems may have distinct optimal electoral periods, a notable question for future research. Second, we find that increasing the ideological sensitivity of candidates reduces robustness, as it increases the chance of overshoot and electoral backlash (Proposition 5). 

We also note limitations of the modeling approach. First, our treatment of the relevant resource (e.g., water) as exogenous does not consider the effects of use on renewable resources (e.g., overuse of groundwater), as done in the \emph{MA model} \cite{Muneepeerakul2017,Muneepeerakul2020}. Second, we consider a one-dimensional policy space (tax setting), but as has been well studied, multidimensional policy spaces experience much more chaotic behavior and various forms of policy cycles \cite{mckelvey_intransitivities_1976}. Third, we aggregate infrastructure into one capacity variable, which does not take into account choices of infrastructure type (e.g., buses or rail). Lastly, we note that real systems often include a complicated blend of redistributive mechanisms (e.g., multiple revenue sources), biases (e.g., multiple perspectives), and response capacities (e.g., elections, public comment, or veto points) \cite{chen_innovative_2022}. We embrace these limitations as opportunities for further research on how political economy affects the robustness of shared infrastructure systems.

Major moments of infrastructure investment (e.g., 2021 Bipartisan Infrastructure Law) may move shared infrastructure systems toward a desired state, but the robustness of such desired states requires careful consideration of the \emph{political economic process} responsible for investment and underlining patterns of inequality. Our modeling approach is a helpful stylized means to capture the basic political-economic dynamics in shared infrastructure systems, and our analysis lays out several propositions to orient future empirical work and institutional design relevant to the infrastructure systems critical to our collective well-being. 

\section{Methods}

\subsection{Infrastructure \& Income}

Infrastructure capacity is converted via a step-wise ramp function, $H(I)$, into a harvesting rate \cite{Muneepeerakul2017}. Shared infrastructure use has a max rate of $h$, and private infrastructure use has a max rate of $h_p$ that when combined with the subsistence rate, $w$ (in transportation, this would be walking) is equal $h$ ($h = h_p + \frac{w}{\tilde{\phi}R}$). We normalize $w$ by the mean earning potential, $\tilde{\phi}$, and the resource level, $R$ (Table \ref{tab:params_dimensionless}). 

Income is the product of $H(I)$, $R$, and the user's earning potential, $\phi_g$. Given their labor allocation choice ($l$), per-user income in group $g$ from shared and private infrastructure, respectively, are given as, 

\begin{align}
    y^s_g &= l_g \phi_g R H^s(I^s) \\
    y^p_g &= (1-l_g)\left(\phi_g RH^p(I^p_g) + w\right)
\end{align}

$Y^s_g$ and $Y^p_g$ denote total shared and private infrastructure-related income in $g$ or $y_gn_g$. $y'_g$ denotes post-tax income or, 

\begin{equation}
    y'g = y^s_g(1-\tau) + y^p_g(1-\psi\tau)
\end{equation} 

Infrastructure capacity increases with investment and depreciates. A coefficient - $\mu$ for shared and $\mu^p_g$ for private infrastructure - converts investments to infrastructure units. We assume all tax revenue is invested. Given depreciation rate $\delta$, shared and private infrastructure capacity follow, 

\begin{align}
    \dot{I}^s &= \mu\tau\left(\sum_g Y^s_g + \psi\sum_g Y^p_g\right) - \delta I^s \\
    \dot{I}^p_g &= \mu^p_g s_g n_g y'_g - \delta I^p_g
\end{align}

\subsection{Labor Allocation \& Savings}

We add private infrastructure to the labor replicator equation from \cite{Muneepeerakul2017}. With sensitivity $\beta^l$, we use,

\begin{multline}
    \dot{l}_g = \beta^l(1-s_g)l_g(1-l_g)[(1-\tau)\phi_gRH^s(I^s)\\-(1-\psi\tau)\left(\phi_gRH(I^p_g) + w\right)] \label{eq:labor}
\end{multline}

We assume savings follow the same decision-making assumptions as labor allocation (derivation in Appendix). Savings will only increase if $I^p_g$ is not at full capacity because further saving would not yield additional income. With sensitivity $\beta^s$, we use,

\begin{equation}
    \dot{s}_g =\beta^s\begin{cases}
        y'_gB(s_g,l_g) & \text{if } \tilde{n}_gI_0 < I^p_g < \tilde{n}_g\bar{I} \\
        -y'_g & \text{otherwise} \label{eq:savings}
    \end{cases}
\end{equation}

where $B(s_g,l_g)$ is, 

\begin{equation}
    B(s_g,l_g) = \frac{\phi_gR(1-\psi\tau)(1-l_g)h_p\mu^p_gN - (\bar{I}-I_0)(1-\delta)}{(\bar{I}-I_0)(1-\delta) - s_g\phi_gR(1-\psi\tau)(1-l_g)h_p\mu^p_gN}
\end{equation}

\subsection{Cold Cognition in DirectAgg Models}

Cold cognition follows a user's observed consumption gradient (derivation in Appendix) consistent with retrospective voter \cite{healy_retrospective_2013,persson_political_2000} and ``cold" Bayesian models \cite{redlawsk_hot_2002}. Given sensitivity $\beta^\tau_1$, users update their preference, $\hat{\tau}_g$, by, 

\begin{equation}
    \dot{\hat{\tau}}_g = \beta^\tau_1 (1-s_g) \begin{cases}
        B^\tau(l_g) - (y^s_g + \psi y^p_g) & \text{if } I_0 < I^s < \bar{I} \\
        - (y^s_g + \psi y^p_g) & \text{otherwise}
    \end{cases} \label{eq:tau_cold}
\end{equation}

where $B^\tau(l_g)$ is, 

\begin{equation}
    B^\tau(l_g) = 
        \frac{\phi_gRl_g(1-\tau)\mu h[\bar{Y}^s + \psi \bar{Y}^p]}{(\bar{I} - I_0) - \mu\tau\tilde{L}Rh}
\end{equation}

and $\tilde{L} = \sum_g \phi_g l_g n_g$. When shared infrastructure is at full capacity, there is no consumption benefit to increasing taxes.  

\subsection{Hot Cognition in DirectAgg Models}

Hot cognition follows a simple motivated reasoning model where voters respond to errors, $e_g$, between perceived and expected consumption and update their tax preference in the direction of their solution bias \cite{schulz_political_2024}, $\theta_g \in \{-1,1\}$. Given sensitivity $\beta^\tau_2$, tax preference, $\hat{\tau}_g$, updates as follows, 

\begin{align}
    \dot{\hat{\tau}}_g &= \beta^\tau_1 \frac{\partial \pi_g}{\partial \tau} + \beta^\tau_2 \theta_g e_g \\
    e_g &= \begin{cases}
        \hat{\pi}_g - \pi_g & \text{if } \hat{\pi}_g > \pi_g \\
        0 & \text{otherwise}
    \end{cases}
\end{align}

One can consider mixed hot and cold cognition with weights $\beta^\tau_1$ and $\beta^\tau_2$ \cite{little_motivated_2022}, but here, we consider one at a time. Expected consumption, $\hat{\pi}_g$, updates with habituation rate, $\xi$, as follows, 

\begin{equation}
    \dot{\hat{\pi}}_g = \xi \left(\pi_g-\hat{\pi}_g\right)
\end{equation}

\subsection{Voter Aggregation in DirectAgg Models}

We convert preferences, $\hat{\tau}_g$, into policy with one of two aggregation rules: \emph{median voter} or \emph{elite capture}. Given that there are more non-elites than elites ($n_2>n_1$), the median voter rule results in non-elite control while elite capture results in elite control. We define $\alpha$ as the aggregation rule parameter where $\alpha=1$ denotes elite capture and $\alpha=0$ denotes median voter, so $\tau$ updates as follows, 

\begin{equation}
    \dot{\tau} = \alpha \dot{\hat{\tau}}_1 + (1-\alpha) \dot{\hat{\tau}}_2
\end{equation}

\subsection{Political Competition}

We consider two candidates, indexed by $q$, who compete for tax-setting authority (the incumbent). We make four assumptions to define political competition in continuous time: (i) every $T_e$ time units, there is an election, (ii) the incumbent can change the tax rate at any time while in office, (iii) candidates are genuine (they enact their platform), and (iv) all voters vote. As discussed above, we assume ideologically motivated candidates, but we define this motivation with a directional bias, $\check{\theta}_q$. Elections are the only signal of voter preferences. Platform, $\check{\tau}_q$, updating is the sum of ideological repulsion with sensitivity, $\sigma^R$, and status quo attraction with sensitivity, $\sigma^A$, or,

\begin{equation}
    \check{\tau}_q = \sigma^A(\tau-\check{\tau}_q) + \sigma^R\check{\theta}_qE(\cdot)
\end{equation}

where $E(\cdot)$ is the total system error, measured by a sum of each group's error, weighted by $J_g$ (defined below).

We impose three constraints on platforms through a threshold function (see Appendix): (i) tax rates shall not be negative, (ii) tax rates shall not exceed available income (excess of subsistence), and (iii) tax rates shall not exceed that which is necessary to maintain full infrastructure capacity, plus a given safety factor, $f_s$ (see Appendix for safety factor sensitivity analysis). 

Following the spatial theory of voter behavior \cite{Kollman2006}, group $g$ will vote for the platform closest to $\hat{\tau}_g$. $L(\hat{\tau}_g,\check{\tau}_q)$ is the quadratic distance between $\check{\tau}_q$ and $\hat{\tau}_g$, so the vote of each group, $v_g$, is,  

\begin{equation}
    v_g = \begin{cases}
        1 & \text{if } L(\hat{\tau}_g,\check{\tau}_1) \leq L(\hat{\tau}_g,\check{\tau}_2) \\
        0 & \text{otherwise}
    \end{cases}
\end{equation}

where $v_g=1$ is voting for candidate 1. $V(\cdot)$ is the number of votes for candidate 1 as a proportion of the population. We consider two electoral conditions: (i) equitable influence (one person, one vote) and (ii) income-based influence (one dollar, one vote). $J_g$ is the relative influence of group $g$, calculated as, 

\begin{equation}
    J_g = \alpha \check{y}_g + (1-\alpha)\tilde{n}_g
\end{equation}

where $\alpha=1$ implies income-based influence, and $\check{y}_g$ is the proportion of total income generated by $g$. $V(\cdot)$ is then,

\begin{equation}
    V(\cdot) = \sum_g \left( v_gJ_g\right)
\end{equation}

At election time, the winning candidate becomes the incumbent, noted by $q^I$, or,     

\begin{equation}
    \dot{q}^I = \begin{cases}
        1 & \text{if } V(\cdot) \geq 0.5 \\
        2 & \text{otherwise}
    \end{cases}
\end{equation}

$q_C$ notes the challenger. Tax policy follows the $q^I$ platform or, 

\begin{equation}
    \dot{\tau} = \check{\tau}_{q^I} - \tau 
\end{equation}

Under \emph{PolComp}, voter preferences move towards platforms based on $e_g$. Success increases the incumbent's pull, $M^I(e_g)$, by, 

\begin{equation}
    M^I(e_g) = \exp\left(-\omega \frac{e_g}{\bar{\pi}_g}\right)
\end{equation}

where $\omega$ is a rate parameter, and $\bar{\pi}_g$ is $g$'s maximum income. The challenger's pull is $1 - M^I(e_g)$, so preferences update as, 

\begin{equation}
    \dot{\hat{\tau}}_g = M^I(e_g)\left(\check{\tau}_{q_I} - \hat{\tau}_g\right) + \left(1-M^I(e_g)\right)\left(\check{\tau}_{q_C} - \hat{\tau}_g\right)  
\end{equation}

\subsection{Model Implementation}

The initial condition is full shared infrastructure ($I^s=\bar{I}$) and nearly full labor allocation in the shared system ($l_g = 0.9$). Elites begin with partial private infrastructure ($I^p_1 = 0.5\tilde{n}_1\bar{I}$) and some saving ($s_1=0.05$). We numerically solve the model for 400 time units with a step of 0.1 using the Tsit5 solver in the DynamicalSystems.jl Julia package \cite{datseris_dynamicalsystemsjl_2018}. Elections in \emph{PolComp} models and shocks in the stochastic analysis were implemented via callbacks in the numerical solver. Based on the average transient time to switch from full shared to full private infrastructure (40 time steps), 400 steps can be conceptualized as 10 possible transition periods.

\section*{Acknowledgements}

  Wiechman acknowledges support from the National Science Foundation Graduate Research Fellowship Program (No. 026257-001), and all authors acknowledge support from the National Science Foundation Grant ``Transition Dynamics in Integrated Urban Water Systems'' (No. 1923880). The views and conclusions expressed here are those of the authors and do not necessarily reflect the views of the National Science Foundation.



\begin{thebibliography}{}

\bibitem[Acemoglu and Johnson, 2023]{acemoglu_power_2023}
Acemoglu, D. and Johnson, S. (2023).
\newblock {\em Power and {Progress}: {Our} {Thousand}-{Year} {Struggle} {Over}
  {Technology}}.
\newblock PublicAffairs.

\bibitem[Acemoglu and Robinson, 2000]{acemoglu_why_2000}
Acemoglu, D. and Robinson, J.~A. (2000).
\newblock Why {Did} the {West} {Extend} the {Franchise}? {Democracy},
  {Inequality}, and {Growth} in {Historical} {Perspective}.
\newblock {\em The Quarterly Journal of Economics}, 115(4):1167--1199.

\bibitem[Acemoglu and Robinson, 2001]{acemoglu_theory_2001}
Acemoglu, D. and Robinson, J.~A. (2001).
\newblock A {Theory} of {Political} {Transitions}.
\newblock {\em American Economic Review}, 91(4):938--963.

\bibitem[Alm, 2015]{alm_financing_2015}
Alm, J. (2015).
\newblock Financing {Urban} {Infrastructure}: {Knowns}, {Unknowns}, and a {Way}
  {Forward}.
\newblock {\em Journal of Economic Surveys}, 29(2):230--262.
\newblock \_eprint: https://onlinelibrary.wiley.com/doi/pdf/10.1111/joes.12045.

\bibitem[An, 2012]{An2012}
An, L. (2012).
\newblock Modeling human decisions in coupled human and natural systems:
  {Review} of agent-based models.
\newblock {\em Ecological Modelling}, 229:25--36.

\bibitem[Anderies, 2015]{Anderies2015b}
Anderies, J.~M. (2015).
\newblock Managing variance: {Key} policy challenges for the {Anthropocene}.
\newblock {\em Proceedings of the National Academy of Sciences of the United
  States of America}, 112(47):14402--14403.

\bibitem[Anderies et~al., 2022]{anderies_framework_2022}
Anderies, J.~M., Cumming, G.~S., Clements, H.~S., Lade, S.~J., Seppelt, R.,
  Chawla, S., and Müller, B. (2022).
\newblock A framework for conceptualizing and modeling social-ecological
  systems for conservation research.
\newblock {\em Biological Conservation}, 275:109769.

\bibitem[Anderies et~al., 2013]{Anderies2013a}
Anderies, J.~M., Folke, C., Walker, B., and Ostrom, E. (2013).
\newblock Aligning key concepts for global change policy: {Robustness},
  resilience, and sustainability.
\newblock {\em Ecology and Society}, 18(2).

\bibitem[Anderies and Janssen, 2013]{Anderies2013b}
Anderies, J.~M. and Janssen, M.~A. (2013).
\newblock Robustness of social-ecological systems: {Implications} for public
  policy.
\newblock {\em Policy Studies Journal}, 41(3):513--536.

\bibitem[Anderies et~al., 2019]{anderies_knowledge_2019}
Anderies, J.~M., Mathias, J.~D., and Janssen, M.~A. (2019).
\newblock Knowledge infrastructure and safe operating spaces in
  social–ecological systems.
\newblock {\em Proceedings of the National Academy of Sciences of the United
  States of America}, 116(12):5277--5284.

\bibitem[Axelrod et~al., 2021]{axelrod_preventing_2021}
Axelrod, R., Daymude, J.~J., and Forrest, S. (2021).
\newblock Preventing extreme polarization of political attitudes.
\newblock {\em Proceedings of the National Academy of Sciences},
  118(50):e2102139118.
\newblock Publisher: Proceedings of the National Academy of Sciences.

\bibitem[Brownlee, 2020]{baker_tax_2020}
Brownlee, W.~E. (2020).
\newblock Tax and {Fiscal} {Regimes} in the {United} {States}: {The} {Long}
  {Swings}.
\newblock In Baker, P. and Critchlow, D.~T., editors, {\em The {Oxford}
  {Handbook} of {American} {Political} {History}}, pages 311--338. Oxford
  University Press.

\bibitem[Chen and Bartle, 2022]{chen_innovative_2022}
Chen, C. and Bartle, J.~R. (2022).
\newblock {\em Innovative {Infrastructure} {Finance}: {A} {Guide} for {State}
  and {Local} {Governments}}.
\newblock Palgrave Macmillan.

\bibitem[Chester et~al., 2020]{chester_infrastructure_2020}
Chester, M.~V., Miller, T., and Muñoz-Erickson, T.~A. (2020).
\newblock Infrastructure governance for the {Anthropocene}.
\newblock {\em Elementa: Science of the Anthropocene}, 8(1):078.

\bibitem[Datseris, 2018]{datseris_dynamicalsystemsjl_2018}
Datseris, G. (2018).
\newblock {DynamicalSystems}.jl: {A} {Julia} software library for chaos and
  nonlinear dynamics.
\newblock {\em Journal of Open Source Software}, 3(23):598.

\bibitem[Davidson et~al., 2024]{davidson_simulating_2024}
Davidson, M.~R., Filatova, T., Peng, W., Verbeek, L., and Kucuksayacigil, F.
  (2024).
\newblock Simulating institutional heterogeneity in sustainability science.
\newblock {\em Proceedings of the National Academy of Sciences},
  121(8):e2215674121.
\newblock Publisher: Proceedings of the National Academy of Sciences.

\bibitem[de~Marchi, 2003]{de_marchi_voter_2003}
de~Marchi, S. (2003).
\newblock Voter {Sophistication}, {Ideology}, and {Candidate} {Position}
  {Taking}.
\newblock In {\em Computational {Models} in {Political} {Economy}}. MIT Press,
  Cambridge, MA.

\bibitem[Deitz and Meehan, 2019]{deitz_plumbing_2019}
Deitz, S. and Meehan, K. (2019).
\newblock Plumbing {Poverty}: {Mapping} {Hot} {Spots} of {Racial} and
  {Geographic} {Inequality} in {U}.{S}. {Household} {Water} {Insecurity}.
\newblock {\em Annals of the American Association of Geographers},
  109(4):1092--1109.

\bibitem[Deslatte et~al., 2022]{deslatte_assessing_2022}
Deslatte, A., Helmke‐Long, L., Anderies, J.~M., Garcia, M., Hornberger,
  G.~M., and Koebele, E. (2022).
\newblock Assessing sustainability through the {Institutional} {Grammar} of
  urban water systems.
\newblock {\em Policy Studies Journal}, 50:387--406.

\bibitem[Downs, 1957]{downs_economic_1957}
Downs, A. (1957).
\newblock {\em An {Economic} {Theory} of {Democracy}}.
\newblock Harper, New York.

\bibitem[Druckman and McGrath, 2019]{druckman_evidence_2019}
Druckman, J.~N. and McGrath, M.~C. (2019).
\newblock The evidence for motivated reasoning in climate change preference
  formation.
\newblock {\em Nature Climate Change}, 9(2):111--119.

\bibitem[Duggan and Martinelli, 2017]{duggan_political_2017}
Duggan, J. and Martinelli, C. (2017).
\newblock The {Political} {Economy} of {Dynamic} {Elections}: {Accountability},
  {Commitment}, and {Responsiveness}.
\newblock {\em Journal of Economic Literature}, 55(3):916--984.

\bibitem[Folke et~al., 2021]{folke_our_2021}
Folke, C., Polasky, S., Rockström, J., Galaz, V., Westley, F., Lamont, M.,
  Scheffer, M., Österblom, H., Carpenter, S.~R., Chapin, F.~S., Seto, K.~C.,
  Weber, E.~U., Crona, B.~I., Daily, G.~C., Dasgupta, P., Gaffney, O., Gordon,
  L.~J., Hoff, H., Levin, S.~A., Lubchenco, J., Steffen, W., and Walker, B.~H.
  (2021).
\newblock Our future in the {Anthropocene} biosphere.
\newblock {\em Ambio}, 50(4):834--869.

\bibitem[Fowler and Smirnov, 2005]{fowler_dynamic_2005}
Fowler, J. and Smirnov, O. (2005).
\newblock Dynamic {Parties} and {Social} {Turnout}: {An} {Agent}‐{Based}
  {Model}.
\newblock {\em American Journal of Sociology}, 110(4):1070--1094.
\newblock Publisher: The University of Chicago Press.

\bibitem[Franzese~Jr., 2002]{franzese_jr_electoral_2002}
Franzese~Jr., R.~J. (2002).
\newblock Electoral and {Partisan} {Cycles} in {Economic} {Policies} and
  {Outcomes}.
\newblock {\em Annual Review of Political Science}, 5:369--421.
\newblock Publisher: Annual Reviews.

\bibitem[Frischmann, 2012]{frischmann_infrastructure_2012}
Frischmann, B.~M. (2012).
\newblock {\em Infrastructure: {The} {Social} {Value} of {Shared} {Resources}}.
\newblock Oxford University Press.

\bibitem[Glaeser and Ponzetto, 2018]{glaeser_political_2018}
Glaeser, E.~L. and Ponzetto, G.~A. (2018).
\newblock The political economy of transportation investment.
\newblock {\em Economics of Transportation}, 13:4--26.

\bibitem[Healy and Malhotra, 2013]{healy_retrospective_2013}
Healy, A. and Malhotra, N. (2013).
\newblock Retrospective {Voting} {Reconsidered}.
\newblock {\em Annual Review of Political Science}, 16(1):285--306.

\bibitem[Helmrich et~al., 2023]{helmrich_lock-_2023}
Helmrich, A., Chester, M., Miller, T.~R., and Allenby, B. (2023).
\newblock Lock-in: origination and significance within infrastructure systems.
\newblock {\em Environmental Research: Infrastructure and Sustainability},
  3(3):032001.
\newblock Publisher: IOP Publishing.

\bibitem[Ingberman, 1989]{ingberman_reputational_1989}
Ingberman, D.~E. (1989).
\newblock Reputational dynamics in spatial competition.
\newblock {\em Mathematical and Computer Modelling}, 12(4):479--496.

\bibitem[Kanoi et~al., 2022]{kanoi_what_2022}
Kanoi, L., Koh, V., Lim, A., Yamada, S., and Dove, M.~R. (2022).
\newblock ‘{What} is infrastructure? {What} does it do?’: anthropological
  perspectives on the workings of infrastructure(s).
\newblock {\em Environmental Research: Infrastructure and Sustainability},
  2(1):012002.
\newblock Publisher: IOP Publishing.

\bibitem[Karabarbounis, 2011]{karabarbounis_one_2011}
Karabarbounis, L. (2011).
\newblock One {Dollar}, {One} {Vote}.
\newblock {\em The Economic Journal}, 121(553):621--651.

\bibitem[Klein, 2025]{klein_opinion_2025}
Klein, E. (2025).
\newblock Opinion {\textbar} {Biden} {Promised} to ‘{Turn} the {Page}’ on
  {Trump}. {What} {Went} {Wrong}?
\newblock {\em The New York Times}.

\bibitem[Kollman et~al., 1992]{kollman_adaptive_1992}
Kollman, K., Miller, J.~H., and Page, S.~E. (1992).
\newblock Adaptive {Parties} in {Spatial} {Elections}.
\newblock {\em The American Political Science Review}, 86(4):929.
\newblock Num Pages: 9 Place: Washington, United Kingdom Publisher: Cambridge
  University Press.

\bibitem[Kollman and Page, 2006]{Kollman2006}
Kollman, K. and Page, S.~E. (2006).
\newblock Chapter 29 {Computational} {Methods} and {Models} of {Politics}.
\newblock {\em Handbook of Computational Economics}, 2(05):1433--1463.
\newblock ISBN: 9780444512536.

\bibitem[Krueger et~al., 2019]{krueger_resilience_2019}
Krueger, E.~H., Borchardt, D., Jawitz, J.~W., Klammler, H., Yang, S., Zischg,
  J., and Rao, P.~S. (2019).
\newblock Resilience {Dynamics} of {Urban} {Water} {Supply} {Security} and
  {Potential} of {Tipping} {Points}.
\newblock {\em Earth's Future}, 7(10):1167--1191.
\newblock Publisher: John Wiley and Sons Inc.

\bibitem[Levin et~al., 2022]{levin_governance_2022}
Levin, S.~A., Anderies, J.~M., Adger, N., Barrett, S., Bennett, E.~M.,
  Cardenas, J.~C., Carpenter, S.~R., Crépin, A.-S., Ehrlich, P., Fischer, J.,
  Folke, C., Kautsky, N., Kling, C., Nyborg, K., Polasky, S., Scheffer, M.,
  Segerson, K., Shogren, J., van~den Bergh, J., Walker, B., Weber, E.~U., and
  Wilen, J. (2022).
\newblock Governance in the {Face} of {Extreme} {Events}: {Lessons} from
  {Evolutionary} {Processes} for {Structuring} {Interventions}, and the {Need}
  to {Go} {Beyond}.
\newblock {\em Ecosystems}, 25(3):697--711.

\bibitem[Levin and Weber, 2024]{levin_polarization_2024}
Levin, S.~A. and Weber, E.~U. (2024).
\newblock Polarization and the {Psychology} of {Collectives}.
\newblock {\em Perspectives on Psychological Science}, 19(2):335--343.
\newblock Publisher: SAGE Publications Inc.

\bibitem[Little et~al., 2022]{little_motivated_2022}
Little, A.~T., Schnakenberg, K.~E., and Turner, I.~R. (2022).
\newblock Motivated {Reasoning} and {Democratic} {Accountability}.
\newblock {\em American Political Science Review}, 116(2):751--767.

\bibitem[McKelvey, 1976]{mckelvey_intransitivities_1976}
McKelvey, R.~D. (1976).
\newblock Intransitivities in multidimensional voting models and some
  implications for agenda control.
\newblock {\em Journal of Economic Theory}, 12(3):472--482.

\bibitem[Mullin and Hansen, 2022]{mullin_local_2022}
Mullin, M. and Hansen, K. (2022).
\newblock Local news and the electoral incentive to invest in infrastructure.
\newblock {\em American Political Science Review}, 117(3):1145--1150.
\newblock ISBN: 0003055422001.

\bibitem[Muneepeerakul and Anderies, 2017]{Muneepeerakul2017}
Muneepeerakul, R. and Anderies, J.~M. (2017).
\newblock Strategic behaviors and governance challenges in social-ecological
  systems.
\newblock {\em Earth's Future}, 5(8):865--876.

\bibitem[Muneepeerakul and Anderies, 2020]{Muneepeerakul2020}
Muneepeerakul, R. and Anderies, J.~M. (2020).
\newblock The emergence and resilience of self-organized governance in coupled
  infrastructure systems.
\newblock {\em Proceedings of the National Academy of Sciences of the United
  States of America}, 117(9):4617--4622.

\bibitem[Pagano, 2011]{pagano_funding_2011}
Pagano, M.~A. (2011).
\newblock Funding and {Investing} in {Infrastructure}.
\newblock Technical report, Urban Institute.

\bibitem[Pandey et~al., 2022]{pandey_infrastructure_2022}
Pandey, B., Brelsford, C., and Seto, K.~C. (2022).
\newblock Infrastructure inequality is a characteristic of urbanization.
\newblock {\em Proceedings of the National Academy of Sciences},
  119(15):e2119890119.
\newblock Publisher: Proceedings of the National Academy of Sciences.

\bibitem[Persson and Tabellini, 2000]{persson_political_2000}
Persson, T. and Tabellini, G. (2000).
\newblock {\em Political {Economics}: {Explaining} {Economic} {Policy}}.
\newblock The MIT Press, Cambridge, MA.

\bibitem[Redlawsk, 2002]{redlawsk_hot_2002}
Redlawsk, D.~P. (2002).
\newblock Hot {Cognition} or {Cool} {Consideration}? {Testing} the {Effects} of
  {Motivated} {Reasoning} on {Political} {Decision} {Making}.
\newblock {\em The Journal of Politics}, 64(4):1021--1044.

\bibitem[Reynolds et~al., 2022]{reynolds_digital_2022}
Reynolds, R., Aromi, J., McGowan, C., and Paris, B. (2022).
\newblock Digital divide, critical-, and crisis-informatics perspectives on
  {K}-12 emergency remote teaching during the pandemic.
\newblock {\em Journal of the Association for Information Science and
  Technology}, 73(12):1665--1680.
\newblock \_eprint: https://onlinelibrary.wiley.com/doi/pdf/10.1002/asi.24654.

\bibitem[Romero-Lankao et~al., 2022]{romero-lankao_inequality_2022}
Romero-Lankao, P., Wilson, A., and Zimny-Schmitt, D. (2022).
\newblock Inequality and the future of electric mobility in 36 {U}.{S}.
  {Cities}: {An} innovative methodology and comparative assessment.
\newblock {\em Energy Research \& Social Science}, 91:102760.

\bibitem[Schulz and Bhui, 2024]{schulz_political_2024}
Schulz, L. and Bhui, R. (2024).
\newblock Political reinforcement learners.
\newblock {\em Trends in Cognitive Sciences}, 28(3):210--222.

\bibitem[Selin et~al., 2023]{selin_progress_2023}
Selin, N.~E., Giang, A., and Clark, W.~C. (2023).
\newblock Progress in modeling dynamic systems for sustainable development.
\newblock {\em Proceedings of the National Academy of Sciences},
  120(40):e2216656120.
\newblock Publisher: Proceedings of the National Academy of Sciences.

\bibitem[Siegel, 2018]{siegel_analyzing_2018}
Siegel, D.~A. (2018).
\newblock Analyzing {Computational} {Models}.
\newblock {\em American Journal of Political Science}, 62(3):745--759.

\bibitem[Sovacool et~al., 2022]{sovacool_towards_2022}
Sovacool, B.~K., Barnacle, M.~L., Smith, A., and Brisbois, M.~C. (2022).
\newblock Towards improved solar energy justice: {Exploring} the complex
  inequities of household adoption of photovoltaic panels.
\newblock {\em Energy Policy}, 164:112868.

\bibitem[Titolo, 2023]{titolo_bidens_2023}
Titolo, M. (2023).
\newblock Biden's {Infrastructure} {Plan}: {A} {New} {Commitment} to {Public}
  {Goods}? {The} {Return} of {Public} {Goods}.
\newblock {\em University of St. Thomas Journal of Law and Public Policy
  (Minnesota)}, 16(1):188--197.

\bibitem[Tocqueville, 1838]{tocqueville_democracy_1838}
Tocqueville, A.~d. (1838).
\newblock {\em Democracy in {America}}.
\newblock Dearborn \& Co., New York.

\bibitem[Vegh and Vuletin, 2015]{vegh_how_2015}
Vegh, C.~A. and Vuletin, G. (2015).
\newblock How {Is} {Tax} {Policy} {Conducted} {Over} the {Business} {Cycle}?
\newblock {\em American Economic Journal: Economic Policy}, 7(3):327--370.
\newblock Publisher: American Economic Association.

\bibitem[Wegrich and Hammerschmid, 2017]{wegrich_infrastructure_2017}
Wegrich, K. and Hammerschmid, G. (2017).
\newblock Infrastructure {Governance} as {Political} {Choice}.
\newblock In Wegrich, K., Kostka, G., and Hammerschmid, G., editors, {\em The
  {Governance} of {Infrastructure}}, pages 21--42. Oxford University Press.

\bibitem[Wiechman et~al., 2024]{wiechman_institutional_2024}
Wiechman, A., Alonso~Vicario, S., Anderies, J.~M., Garcia, M., Azizi, K., and
  Hornberger, G. (2024).
\newblock Institutional {Dynamics} {Impact} the {Response} of {Urban}
  {Socio}-{Hydrologic} {Systems} to {Supply} {Challenges}.
\newblock {\em Water Resources Research}, 60(2):e2023WR035565.

\bibitem[Wittman, 1983]{wittman_candidate_1983}
Wittman, D. (1983).
\newblock Candidate {Motivation}: {A} {Synthesis} of {Alternative} {Theories}.
\newblock {\em American Political Science Review}, 77(1):142--157.

\bibitem[Zheng and Walsham, 2021]{zheng_inequality_2021}
Zheng, Y. and Walsham, G. (2021).
\newblock Inequality of what? {An} intersectional approach to digital
  inequality under {Covid}-19.
\newblock {\em Information and Organization}, 31(1):100341.

\end{thebibliography}

\begin{appendix}

\subsection*{Private and Shared Infrastructure Harvest Functions} 
Infrastructure units are converted via a harvest function, $H(I)$, into a harvesting rate. We follow the MA model example of the step-wise ramp function for $H(I)$, which takes three parameters: a minimum infrastructure level ($I_0$), a maximum infrastructure level ($\bar{I}$), and a maximum harvesting rate ($h$) \cite{Muneepeerakul2017}. To account for the fact that private infrastructure is group-specific (i.e., only group $g$ has access to $I^p_g$, but all groups can access $I^s$), we modify the minimum and maximum infrastructure levels by the fraction of the population contained in $g$, $\tilde{n}_g = \frac{n_g}{N}$. The harvest functions for shared, $H^s(I^s)$, and private infrastructure, $H^p(I^p_g)$, are, thus, defined, respectively, as, 

\begin{align}
      H^s(I^s) &= \begin{cases}
        h & \text{if } I^s \geq \bar{I} \\
        h\frac{I^s - I_0}{\bar{I}-I_0} & \text{if } \bar{I} > I^s > I_0 \\
        0 & \text{otherwise}
    \end{cases} \\
     H^p(I^p_g) &= \begin{cases}
        h_p & \text{if } I^p_g \geq \tilde{n}_g\bar{I} \\
        h_p\frac{I^p_g - \tilde{n}_gI_0}{\tilde{n}_g(\bar{I}-I_0)} & \text{if } \tilde{n}_g\bar{I} > I^p_g > \tilde{n}_g I_0 \\
        0 & \text{otherwise}
    \end{cases}
\end{align}

where $h_p$ is the maximum harvesting rate for private infrastructure. Because private infrastructure augments that which can be earned privately without infrastructure ($w$), we make the assumption that at maximum private infrastructure, one earns the same as they would under maximum shared infrastructure. This allows us to set $h_p = h(1 - \tilde{w})$ where $\tilde{w}$ is the possible private wage (without infrastructure) normalized by $\phi_g$, $R$, and $h$ (inverse of $\phi_1$ variable in \cite{Muneepeerakul2017}).

\subsection*{Derivation of Labor Allocation Dynamics}

The consumption gradient with respect to labor allocation for group $g$, $\frac{\partial \pi_g}{l_g}$, can be re-written in terms of post-tax income as follows, 

\begin{equation}
    \frac{\partial \pi_g}{l_g} = (1-s_g)\frac{\partial y'_g}{l_g} 
\end{equation}

Post-tax income per user is defined as follows (see manuscript Section 3 for explanation), 

\begin{equation}
    y'_g = (1-\tau)l_g\phi_gRH(I^s) + (1-\psi\tau)(1-l_g)\left[\phi_g R H(I^p_g) + w\right]\\
\end{equation}

The partial derivative, then, for post-tax income with respect to labor is as follows, 

\begin{equation}
    \frac{\partial y'_g}{l_g} = (1-\tau)\phi_gRH^s(I^s) - (1-\psi\tau)\left[H(I^p_g)\phi_gR + w\right] + (1-\tau)l_g\phi_gR\frac{\partial H^s(I^s)}{l_g} + (1-\psi\tau)(1-l_g)\phi_gR\frac{\partial H^p(I^p_g)}{l_g}
\end{equation}

Muneepeerakul and Anderies (2017) utilize a replicator equation representation of labor allocation dynamics that assumes users increase or decrease their use of shared infrastructure based on how well the infrastructure performs relative to outside opportunity at the moment of choice and the prevalence of shared infrastructure use (if more people use, more people will want to use). Under this assumption, they do not make predictions about how the lucrativeness might change given their use change. Thus, in developing our consumption gradient for labor allocation, we also assume that the user does not know how the infrastructure productivity will change as they increase use. This implies that, from the user's perspective, the partial derivative of $H^s(I^s)$ and $H^p(I^p_g)$ with respect to $l_g$ is 0, leaving the following equation for $\frac{\partial y'_g}{l_g}$,

\begin{equation}
    \frac{\partial y'_g}{l_g} = (1-\tau)\phi_gRH^s(I^s) - (1-\psi\tau)\left[H(I^p_g)\phi_gR + w\right] \label{eq:posttax_y}
\end{equation}

Substituting this equation into the above equation for $\frac{\partial \pi_g}{l_g}$ yields, 

\begin{equation}
    \frac{\partial \pi_g}{l_g} = (1-s_g)\left((1-\tau)\phi_gRH^s(I^s) - (1-\psi\tau)\left[H(I^p_g)\phi_gR + w\right]\right)
\end{equation}

The only difference between this equation, and the one used by Muneepeerakul and Anderies (2017), besides the fact that they do not account for savings or private infrastructure, is the $l(1-l)$ term. We decide to add the $l_g(1-l_g)$ term to the expression for $\dot{l}$ to accommodate the Muneepeerakul and Anderies (2017) assumption that labor allocation decisions follow a replicator relationship where more use begets more use ($l$ term), but there are limits to use ($1-l$ term). Taken together, we arrive at the labor allocation dynamics used in the manuscript (Manuscript Equation 7).

\subsection*{Derivation of Savings Rate Dynamics}

Like labor allocation, we begin with $\frac{\partial \pi_g}{s_g}$, which is, by the product rule, 

\begin{equation}
    \frac{\partial \pi_g}{\partial s_g} = (1-s_g)\frac{\partial y'_g}{\partial s_g} - y'_g \label{eq:dpids_simple}
\end{equation}

With reference to Equation \ref{eq:posttax_y}, $\frac{\partial y'_g}{s_g}$ is, 

\begin{equation}
    \frac{\partial y'_g}{\partial s_g} = (1-\tau)\phi_gR\frac{\partial H^s(I^s)}{\partial s_g} + (1-\psi\tau)(1-l_g)\phi_gR\frac{\partial H^p(I^p_g)}{\partial s_g} \label{eq:dypostds}
\end{equation}

Like labor allocation, we assume boundedly rational users who do not  know how shared infrastructure productivity ($H^s(I^s)$) will change as they increase savings ($\frac{\partial H^s(I^s)}{\partial s_g} = 0$). However, because we assume users save to increase their future private infrastructure-related income, we assume that users know how marginal increases in savings will affect their private infrastructure income, holding all else constant at the current value (or $\frac{\partial H^p(I^p_g)}{\partial s_g}$). $\frac{\partial H^p(I^p_g)}{\partial s_g}$ is non-zero only when private infrastructure is at partial capacity ($\tilde{n}_gI_0<I^p_g<\tilde{n}_g\bar{I}$). It is given as, 

\begin{equation}
    \frac{\partial H^p(I^p_g)}{\partial s_g} = \begin{cases}
        \frac{h_p}{\tilde{n}_g(\bar{I}-I_0)} \frac{\partial I^p_g}{\partial s_g} & \text{if } \tilde{n}_gI_0<I^p_g<\tilde{n}_g\bar{I} \\
        0 & \text{otherwise} 
    \end{cases} \label{eq:dHpds}
\end{equation}

By the product rule, $\frac{\partial I^p_g}{\partial s_g}$ is given as, 

\begin{equation}
    \frac{\partial I^p_g}{\partial s_g} = \mu^p_gn_g\left[y'_g + s_g \frac{\partial y'_g}{\partial s_g}\right] \label{eq:dIpds}
\end{equation}

Substituting Equations \ref{eq:dHpds} and \ref{eq:dIpds} into Equation \ref{eq:dypostds} and re-arranging with the definition, $\tilde{n}_g = \frac{n_g}{N}$, yields, 

\begin{equation}
    \frac{\partial y'_g}{\partial s_g} = 
    \begin{cases}
        y'_g \frac{(1-\psi\tau)(1-l_g)\phi_gRh_p\mu^p_gN}{(\bar{I}-I_0) - s_g(1-\psi\tau)(1-l_g)\phi_gRh_p\mu^p_gN} & \text{if } \tilde{n}_gI_0 < I^p_g < \tilde{n}_g\bar{I} \\
        0 & \text{otherwise}
    \end{cases}
    \label{eq:dypostds_long}
\end{equation}

Substituting Equation \ref{eq:dypostds_long} into Equation \ref{eq:dpids_simple} and re-arranging yields,

\begin{equation}
   \frac{\partial \pi_g}{\partial s_g} = \begin{cases}
        y'_g\left(\frac{\phi_gR(1-\psi\tau)(1-l_g)h_p\mu^p_gN - (\bar{I}-I_0)(1-\delta)}{(\bar{I}-I_0)(1-\delta) - s_g\phi_gR(1-\psi\tau)(1-l_g)h_p\mu^p_gN}\right) & \text{if } \tilde{n}_gI_0 < I^p_g < \tilde{n}_g\bar{I} \\
        -y'_g & \text{otherwise} \label{eq:savings1}
    \end{cases}
\end{equation}

This the gradient, when combined with the sensitivity parameter and the boundaries discussed in Section 3 of the manuscript, applied to the dynamics of the savings rate (Manuscript Equation 8). 

\subsection*{Derivation of Tax Rate Dynamics (Cold Cognition)}

Like savings and labor allocation, we begin with the consumption gradient with respect to the tax rate, $\frac{\partial \pi_g}{\partial \tau}$, given as, 

\begin{equation}
    \frac{\partial \pi_g}{\partial \tau} = (1-s_g)\frac{\partial y'_g}{\partial \tau} \label{eq:dpidtau_simple}
\end{equation}

The post-tax income partial derivative, with the product rule, is given as, 

\begin{equation}
    \frac{\partial y'_g}{\partial \tau} = \phi_gRl_g(1-\tau)\frac{\partial H^s(I^s)}{\partial \tau}-y^s_g + (1-\psi\tau)(1-l_g)\phi_gR\frac{\partial H^p(I^p_g)}{\partial \tau} - \psi y^p_g \label{eq:dyposttax_dtau}
\end{equation}

We assume, for the ``cold" cognition learning (see manuscript for theoretical basis), a similar assumption of user knowledge from that of the savings rate where users have knowledge of how marginal increases in the tax rate will affect shared infrastructure productivity. However, like savings rate, we assume that they are not away of indirect effects on the other infrastructure type ($\frac{\partial H^p(I^p_g)}{\partial \tau}=0$). $\frac{\partial H^s(I^s)}{\partial \tau}$ is given as, 

\begin{equation}
    \frac{\partial H^s(I^s)}{\partial \tau} = \begin{cases}
        \frac{h}{\bar{I}-I_0} \frac{\partial I^s}{\partial \tau} & \text{if } I_0<I^s<\bar{I} \\
        0 & \text{otherwise} 
    \end{cases} \label{eq:dHsdtau}
\end{equation}

Then, $\frac{\partial I^s}{\partial \tau}$ is given as, 

\begin{equation}
    \frac{\partial I^s}{\partial \tau} = \mu\left[\bar{Y}^s + \psi \bar{Y}^p + \tau\left(\frac{\partial \bar{Y}^s}{\partial \tau} + \psi\frac{\partial \bar{Y}^p}{\partial \tau}\right)\right] \label{eq:dIsdtau}
\end{equation}

where $\bar{Y}^s=\sum_g Y^s_g$ and $\bar{Y}^p=\sum_g Y^p_g$. Again, we assume that the user is not aware of how private infrastructure productivity is affected by tax increases, so $\frac{\partial \bar{Y}^p}{\partial \tau} = 0$. However, $\frac{\partial \bar{Y}^s}{\partial \tau}$ is given as, 

\begin{equation}
    \frac{\partial \bar{Y}^s}{\partial \tau} = \tilde{L}R \frac{\partial H^s(I^s)}{\partial \tau}
\end{equation}

Substituting this Equation into Equation \ref{eq:dIsdtau} and \ref{eq:dHsdtau} and re-arranging yields, 

\begin{equation}
    \frac{\partial H^s(I^s)}{\partial \tau} = \begin{cases}
        \frac{h\mu\left(\bar{Y}^s + \bar{Y}^p\right)}{\bar{I}-I_0 - \mu\tau\tilde{L}Rh} & \text{if } I_0<I^s<\bar{I} \\
        0 & \text{otherwise} 
    \end{cases} \label{eq:dHsdtau1}
\end{equation}

Substituting this equation into Equation \ref{eq:dyposttax_dtau} yields, 

\begin{equation}
    \frac{\partial y'_g}{\partial \tau} = 
    \begin{cases}
        \phi_gRl_g(1-\tau)\frac{h\mu\left(\bar{Y}^s + \bar{Y}^p\right)}{\bar{I}-I_0 - \mu\tau\tilde{L}Rh}-(y^s_g + \psi y^p_g) & \text{if } I_0<I^s<\bar{I} \\
        -(y^s_g + \psi y^p_g) & \text{otherwise} 
    \end{cases}\label{eq:dyposttax_dtau1}
\end{equation}

Substituting this equation into Equation \ref{eq:dpidtau_simple} yields,

\begin{equation}
    \frac{\partial \pi_g}{\partial \tau} = (1-s_g)
    \begin{cases}
        \phi_gRl_g(1-\tau)\frac{h\mu\left(\bar{Y}^s + \bar{Y}^p\right)}{\bar{I}-I_0 - \mu\tau\tilde{L}Rh}-(y^s_g + \psi y^p_g) & \text{if } I_0<I^s<\bar{I} \\
        -(y^s_g + \psi y^p_g) & \text{otherwise} 
    \end{cases}\label{eq:dyposttax_dtau2}
\end{equation}

We combine this partial derivative with the boundaries placed on the tax rate and sensitivity parameter discussed in the manuscript to get the cold cognition tax rate preference updating dynamics used in the manuscript (Manuscript Equation 11).

\subsection*{Subsistence Income and Thresholds in Savings and Tax Policy}
We set a level of income needed for each group to satisfy their basic needs (subsistence), $y^0_g$. Users will never save their subsistence income, and we assume policymakers know this level and will thus, not set tax rates that require users to pay their subsistence income in taxes. For savings, a threshold function $T^s(\dot{s}_g,s_g)$ provides the savings rate adopted by a user in group $g$, taking into account their $y^0_g$ and preventing the savings rate from going below 0, and is given as,

\begin{equation}
    T^s(\dot{s}_g,s_g) = \begin{cases}
        s^A(y'_g) - s_g & \text{if } \dot{s}_g > s^A(y'_g) - s_g \\
        -s_g & \text{if } \dot{s}_g < - s_g \\
        \dot{s}_g & \text{otherwise}
    \end{cases}
\end{equation}

where $s^A(y'_g)$ is the proportion of their post-tax income that is available to be saved given $y^0_g$, and is defined as, 

\begin{equation}
    s^A(y'_g) = \frac{y'_g - y^0_g}{y'_g}
\end{equation}

For taxes, we define a threshold function $T^\tau(\dot{\tau},\tau)$ to keep taxes above 0 and below the subsistence level of the non-elites ($\bar{\tau}_1$) and the level needed to reach the maximum desired infrastructure level ($\bar{\tau}_2$). $\bar{\tau}_1$ is a function of current income and $y^0_2$ because group 2 has a lower earning potential ($\bar{\tau}_1 = \frac{y_2 - y^0_2}{y_2}$). $\bar{\tau}_2$ assumes $\bar{I}(1+f_s)$ is the maximum desired infrastructure capacity, as explained in the manuscript, so it is defined as,

\begin{equation}
    \bar{\tau}_2 = \frac{\bar{I}(1+f_s) - (1-\delta)I_s}{\mu(Y_s + \psi Y_p)}
\end{equation}

Together, we define $T^\tau(\dot{\tau},\tau)$ as,

\begin{equation}
    T^\tau(\dot{\tau},\tau) = \begin{cases}
        -\tau & \text{if } \dot{\tau} < -\tau \\
        \bar{\tau}_1 - \tau & \text{if } \dot{\tau} > \bar{\tau}_1 - \tau \quad \& \quad \bar{\tau}_1 < \bar{\tau}_2 \\
        \bar{\tau}_2 - \tau & \text{if } \dot{\tau} > \bar{\tau}_2 - \tau \quad \& \quad \bar{\tau}_1 > \bar{\tau}_2 \\
        \dot{\tau} & \text{otherwise}
    \end{cases} \label{eq:tau_max}
\end{equation}

The savings rate and tax dynamics follow the dynamics discussed in the main text, but in the dynamical systems definition, we incorporate these threshold functions to provide the boundary conditions of the $s_g$ and $\tau$.

\begin{figure}[!t]
    \centering
    \includegraphics[width=\linewidth]{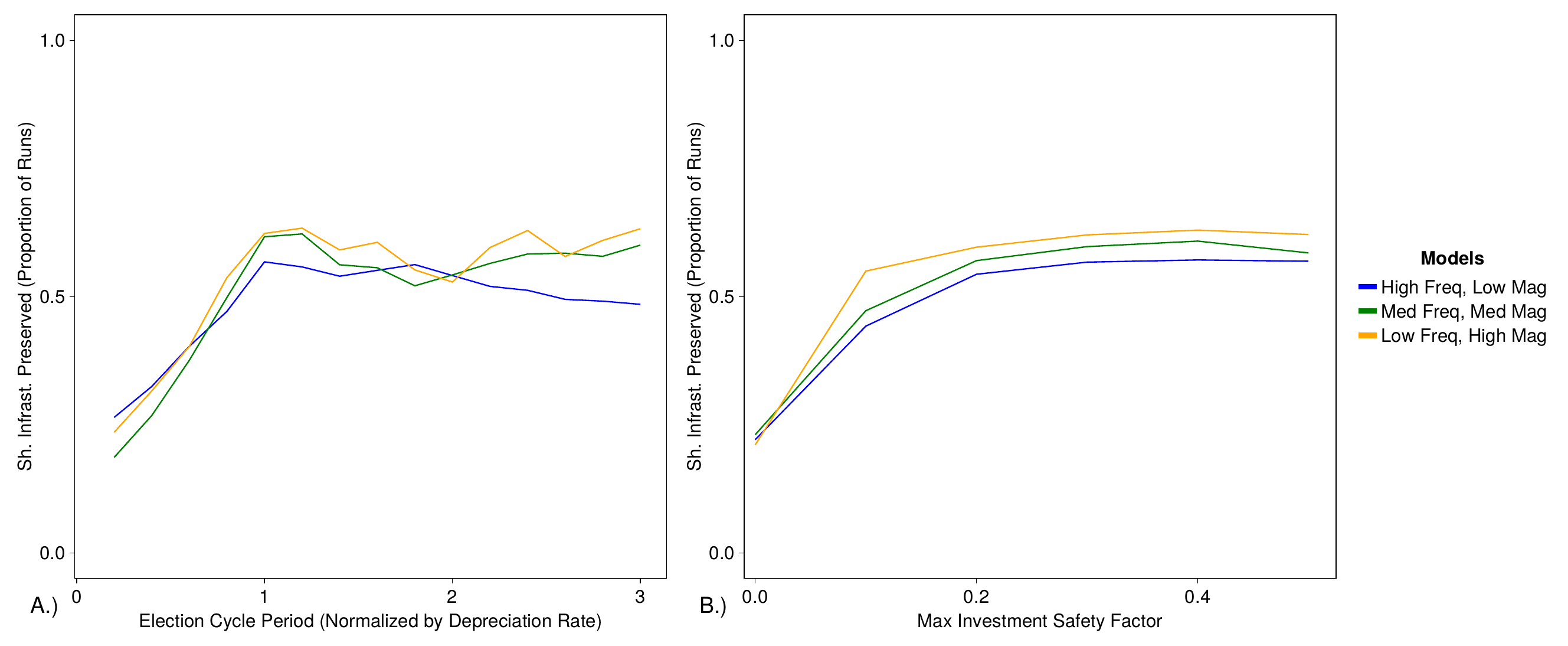}
    \caption{In the stochastic analysis, we allow tax platforms offered by candidates to go beyond $\bar{I}$ (see Subsistence Income and Thresholds in Savings and Tax Policy) to prepare for future shocks given a safety factor ($f_s$). We conducted a two-dimensional sensitivity analysis, varying the election cycle period ($T_e$) and $f_s$, for three representative shock regimes, similar to the response capacity sensitivity analysis shown in the manuscript (Figure \ref{fig:MS_response}) The three shock regimes are high frequency-low magnitude ($T_s$ = 8, $a=0.1$), medium frequency-medium magnitude ($T_s=50$, $a=0.25$), and high frequency-low magnitude ($T_s=150$, $a=0.5$). Similar to the sensitivity analysis reported in the manuscript (Figure \ref{fig:MS_response}), below a normalized election cycle period of 1, increasing $T_e$ benefits robustness (A), but the effect of continuing to increase $T_e$ depends on the shock regime. On the other hand, increasing $f_s$ up until 0.2 (default used in the stochastic analysis) benefits robustness, but beyond 0.2, there are no significant benefits to robustness.} 
    \label{fig:MS_SFEffect}
\end{figure}

\newpage

\begin{figure}[!t]
    \centering
    \includegraphics[width=\linewidth]{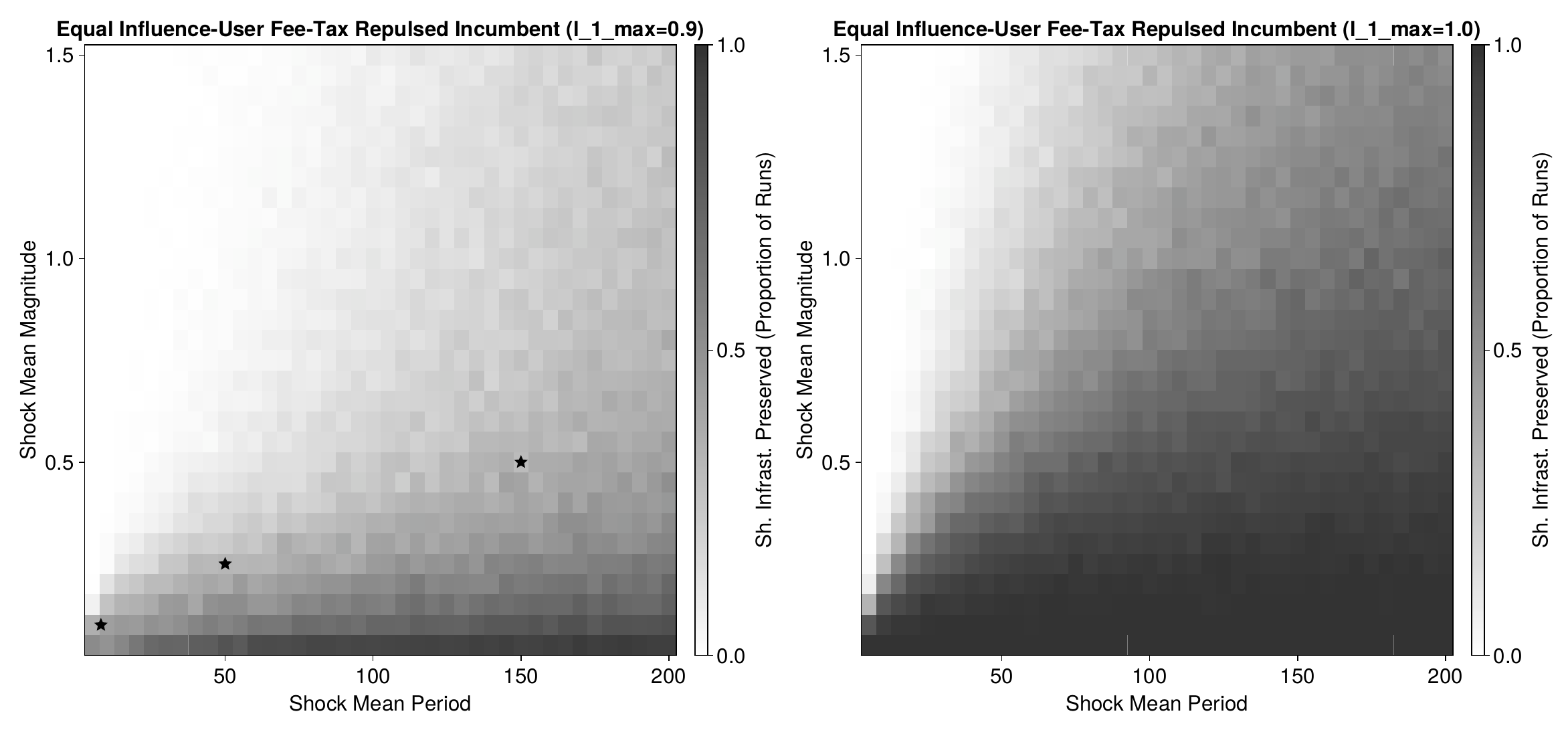}
    \caption{In the stochastic analysis, to ensure that each successive shock in a model run has the potential to cause elite exiting, a cap on elite shared system labor allocation ($\bar{l}_1$) was required. We set the cap to 0.9 to match the initial condition used in the deterministic analysis. This captures the increased vulnerability of the shared system to elite exiting when there are shocks to shared infrastructure capacity (left) compared to a system where the elites drift fully into the shared system ($l_1$ approaches 1) and have no exiting option (right). We note the three example shock regimes with stars used in the stochastic analysis described in the manuscript.} 
    \label{fig:MS_eliteEffect}
\end{figure}

\begin{figure}
    \centering
    \includegraphics[width=\linewidth]{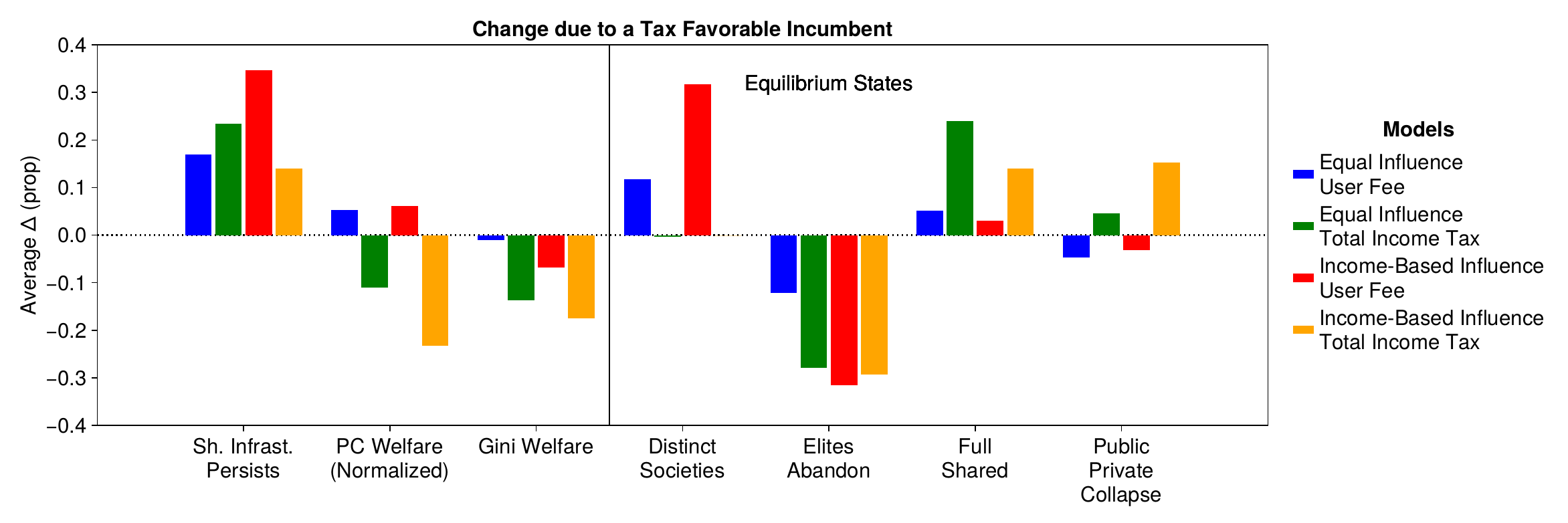}
    \caption{Having an initial incumbent that is tax favorable substantially increases the robustness of the shared infrastructure system. We plot the proprotional average effect to robustness (shared infrastructure persisting), per capita welfare, and gini welfare of switching from a tax-repulsed to tax-favorable incumbent across the other four dimensions of the \emph{PolComp} models. We note the change to post-shock equilibrium states as an explanation for the robustness and welfare outcomes.}
    \label{fig:incumb}
\end{figure}

    \end{appendix}

\end{document}